\pdfoutput=1 
\documentclass[useAMS,usenatbib,onecolumn]{mn2e}
\usepackage{psfig}  
\usepackage{graphicx} 
\usepackage{dcolumn}  
\usepackage{bm}       
\usepackage{amssymb,amsmath}  
\usepackage{color}
\usepackage{float}
\newcommand{\beq}{\begin{equation}}
\newcommand{\eeq}{\end{equation}}

\newcommand{\beqa}{\begin{eqnarray}}

\newcommand{\etal}{{\it et al.}}

\def\Kpc{\, h^{-1} \, {\rm Kpc}}
\def\Mpc{\, h^{-1} \, {\rm Mpc}}
\def\Gpc{\, h^{-1} \, {\rm Gpc}}

\def\kvecMpc{\, h \, {\rm Mpc}^{-1}}
\def\Msun{\,h^{-1}\,{\rm M_{\odot}}}
\def\ltsima{$\; \buildrel < \over \sim \;$}   
\def\gtsima{$\; \buildrel > \over \sim \;$}   
\def\simlt{\lower.5ex\hbox{\ltsima}}   
\def\simgt{\lower.5ex\hbox{\gtsima}}   
\def\etal{{et al. }}

\newcommand{\nc}{\newcommand}   

\nc{\de}{\delta}
\nc{\hn}{\hat{n}}
\nc{\bH}{\bar{H}}
\nc{\Ol}{\Om_{\Lambda}}

\nc{\ul}{\underline} \nc{\al}{\alpha} \nc{\g}{\gamma}
\nc{\Del}{\Delta} \nc{\e}{\textrm{e}} \nc{\eps}{\epsilon}
\nc{\lam}{\lambda} \nc{\Om}{\Omega} \nc{\Omm}{\Omega_m}
\nc{\Oml}{\Omega_\Lambda} \nc{\LCDM}{$\Lambda$CDM~} 
\nc{\ve}{\varepsilon} \nc{\mn}{{\mu\nu}} \nc{\vp}{\varphi}

\def\gsim{\; \raise0.3ex\hbox{$>$\kern-0.75em
\raise-1.1ex\hbox{$\sim$}}\; }

\nc{\Section}[2]{\section{#2}\label{#1}}   
\nc{\Bibitem}[1]{\bibitem{#1}}   
\nc{\Label}[1]{\label{#1}}   

\nc{\hq}{\hat{q}}
\nc{\hw}{\widehat{w}}

\def\ben{\begin{enumerate}}
\def\een{\end{enumerate}}
\def\bi{\begin{itemize}}
\def\ei{\end{itemize}}
\def\ee{\end{equation}}
\def\bea{\begin{eqnarray}}
\def\eea{\end{eqnarray}}

\nc{\M}{\rm{M}}
\nc{\Gpcc}{\rm{~ Gpc^3/h^3}}     
   
\def\etal{{et al. }}   
   
\def\ltsima{$\; \buildrel < \over \sim \;$}   
\def\gtsima{$\; \buildrel > \over \sim \;$}   
\def\simlt{\lower.5ex\hbox{\ltsima}}   
\def\simgt{\lower.5ex\hbox{\gtsima}}   
\nc{\w}{$w(\theta)$\ }   
\nc{\ie}{i.e., }    
\nc{\eg}{e.g., }

\def\Cl{C_{\ell}}

\input epsf

\begin{document}   
   
\title[The MICE Grand Challenge]
{The MICE Grand Challenge Lightcone Simulation I:\\ Dark matter clustering}

\author[Fosalba \etal]{
 P. Fosalba$^{\star}$,  M. Crocce, E. Gazta\~{n}aga, \& F. J. Castander  \\
Institut de Ci\`encies de l'Espai, IEEC-CSIC, Campus UAB, Facultat de
Ci\`encies, Torre C5 par-2, Barcelona 08193, Spain \\ 
\newauthor  
{\footnotesize $^{\star}$ Dedicated to my mother, Florencia} }

\twocolumn   
\maketitle 

\begin{abstract}

We present a new N-body simulation from the MICE collaboration, {\it the
MICE Grand Challenge} (MICE-GC), containing about 70 billion
dark-matter particles in a $(3 Gpc/h)^3$ comoving volume. 
Given its large volume and fine spatial resolution, spanning over 5 orders of magnitude
in dynamic range, it allows an accurate modeling of the growth
of structure in the universe from the linear through the highly non-linear
regime of gravitational clustering. We validate the dark-matter simulation outputs using 3D and 2D 
clustering statistics, and discuss mass-resolution effects in the non-linear
regime by comparing to previous simulations and the latest numerical fits.
We show that the MICE-GC run allows for a measurement of the BAO
feature with percent level accuracy and compare it to
state-of-the-art theoretical models. We also use sub-arcmin
resolution pixelized 2D maps of the dark-matter counts in the
lightcone to make tomographic analyses in real and redshift space. 
Our analysis shows the simulation reproduces the Kaiser effect
on large scales, whereas we find a significant suppression of power on
non-linear scales relative to the real space clustering. 
We complete our validation by presenting an analysis of
the 3-point correlation function in this and previous MICE
simulations, finding further evidence for mass-resolution effects.
This is the first of a series of three papers in which we present the
MICE-GC simulation, along with a wide and deep mock galaxy catalog
built from it. This mock is made publicly available through a dedicated webportal,
{\texttt http://cosmohub.pic.es}.
 
\end{abstract}   
   
\begin{keywords}
methods: numerical,(cosmology):large-scale structure of Universe, (cosmology):dark
matter,galaxies: statistics
\end{keywords}

\section{Introduction}   

These are exciting times for cosmology.
The new generation of astronomical surveys will deliver a detailed
picture of the universe, providing a better understanding of the
galaxy formation process
and determining whether General Relativity correctly describes the
observables that characterize the expansion rate of the universe
and the growth of large-scale structures. In particular one of the
main science goals of upcoming galaxy surveys
is to pin down the properties of the dark-energy that drives
the observed accelerated expansion of the universe.

\begin{figure*}
\begin{center}
\includegraphics[width=0.98\textwidth]{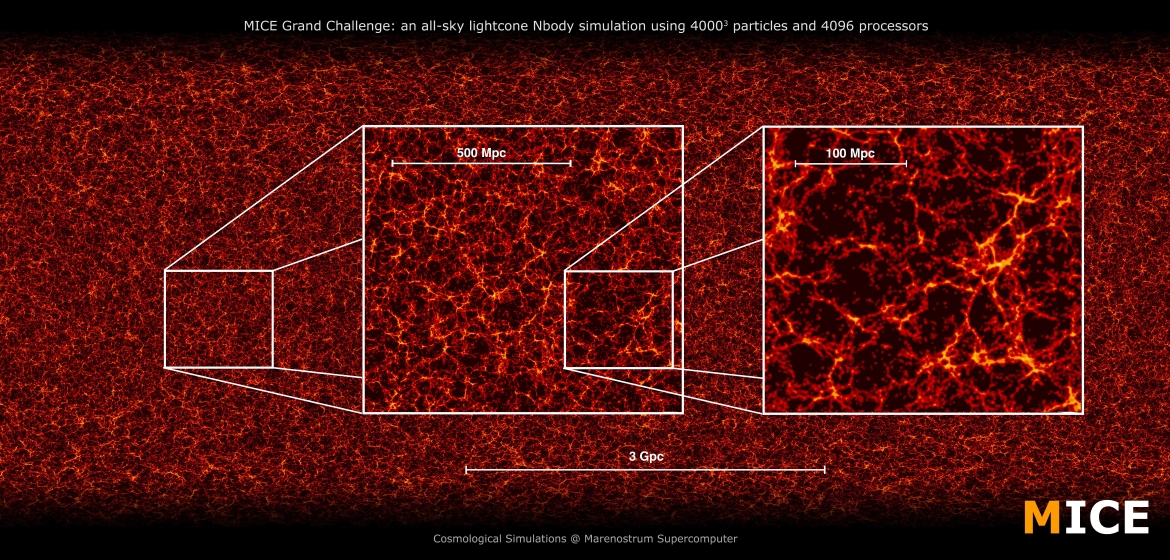} 
\caption{MICE-GC dark-matter lightcone simulation at $z=0.6$.
The image shows the wide dynamic range, about 5 decades in scale, sampled by this Nbody simulation.} 
\label{fig:micesphere}
\end{center}
\end{figure*}

In order to answer these key questions we need to interpret 
astronomical datasets with high precision. Given the difficulty of the
problem, N-body simulations have become
a fundamental ingredient to develop mock cosmological observations, by providing 
a laboratory for the growth of structure under the only effect of gravity.
However, the ever increasing 
volume and complexity of observational datasets demands a matching
effort to develop larger and higher accuracy cosmological simulations.

In this paper we present a new N-body simulation developed by the
Marenostrum Institut de Ci\`encies de l'Espai (MICE)
collaboration at the Marenostrum supercomputer, the MICE {\it Grand
Challenge} run (MICE-GC), that includes about 70 billion dark-matter
particles, in a box of about $3 \Gpc$ aside. This simulation samples from the largest (linear) scales
accessible to the observable universe, where clustering statistics are Gaussian, down to 
to the highly non-linear regime of structure formation where gravity
drives dark-matter and galaxy clustering away from Gaussianity.
We build halo and galaxy catalogues in order to help the design and
exploitation of
new wide-area cosmological surveys. 
We present several applications of this large cosmological simulation 
for 2D and 3D clustering statistics of dark-matter in comoving outputs and
in the lightcone. 

One of the main focus of this paper is to 
investigate the  impact of {\it mass-resolution  effects} 
in the modeling of dark-matter
and galaxy clustering observables by comparing the MICE-GC, and
previous MICE runs, to analytic fits available based on
high-resolution N-body simulations.
For this purpose, we have throughly analyzed the simulation outputs
using basic 3D and 2D clustering statistics in comoving outputs and in
the lightcone. We show that our results are
consistent with previous work in linear and
weakly non-linear scales, and show
how the new simulation better samples power on small-scales, described
by the highly non-linear regime of gravitational clustering.

This paper is the first of a series of three introducing the
MICE-GC end-to-end simulation, its validation and several applications. 
Paper I (i.e. this paper) presents the MICE-GC N-body lightcone simulation and its
validation using dark-matter clustering statistics.
The halo and mock galaxy catalog, along with their validation 
and applications to abundance and clustering statistics are presented in Paper
II \citep{MICE2}. The all-sky lensing maps and the inclusion of
 lensing properties to the mock galaxies are discussed in Paper III \citep{MICE3}.

Accompanying this series of papers, we make a first public data
release of the MICE-GC lightcone galaxy mock ({\tt MICECAT v1.0}) 
through a dedicated web-portal for simulations: 
{\texttt http://cosmohub.pic.es},
where detailed information on the data provided can be found.
We plan to release improved versions of the MICE galaxy mocks through
this web-portal in due time.

This paper is organized as follows: 
Section \S\ref{sec:sim} presents the MICE-GC simulation and how it
compares to state-of-the-art in the field of numerical simulations in cosmology. 
In \S\ref{sec:3dclustering}, we validate the dark-matter outputs of the
simulation using the 3D matter power-spectrum, 
and its Fourier transform, the 2-point correlation function (\ie
2PCF). In \S\ref{sec:cls} we investigate the distribution of dark-matter
in the all-sky lightcone by means of 2D pixelized maps. We focus
in the clustering of
projected dark-matter counts in redshift bins
in real and redshift space using the angular power spectrum
and its Legendre Transform, the angular 2PCF. 
We compare these results to previous simulations and
available numerical fits.
Section \S\ref{sec:3pcf} presents an analysis of the 3-point correlation function of
the dark-matter comoving outputs in the MICE-GC, and compares it to
previous MICE runs and theory predictions.
Finally, in \S\ref{sec:conclusions} we summarize our main results and conclusions.

\begin{table*} 
\begin{center}
\begin{tabular}{lcccccccccccccccc}
\hline \\
Run        &&&    $N_{{\rm part}}$ & \ $L_{{\rm box}}/\Mpc$  \ & \ $PMGrid$ &
$m_p/(10^{10} \Msun)$  & $l_{{\rm soft}}/\Kpc$ & $z_{{\rm i}}$   & $Max. TimeStep$ \\

           &&&  &  & &  & & &    \\

MICE-GC   &&&    $4096^3$  & $3072$    & $4096$    & $2.93$  & $50$  & $100$ & $0.02$       \\

           &&&  &  & &  & & &    \\

MICE-IR   &&&    $2048^3$  & $3072$     & $2048$   & $ 23.42$ &  $50$ & $50$   & $0.01$    \\ 

MICE-SHV   &&&    $2048^3$  & $7680$    & $2048$    & $366$   & $50$ & $150$   & $0.03$ \\  \\

\hline
\end{tabular}
\end{center}
\caption{Description of the MICE N-body simulations. $N_{{\rm part}}$
  denotes number of particles, $L_{{\rm box}}$ is the box-size, $PM
  Grid$ gives the size of the Particle-Mesh grid used for the
  large-scale forces computed with FFTs, $m_p$ gives the particle mass, $l_{soft}$ is the softening length,
and  $z_{in}$ is the initial redshift of the simulation. All
simulations had initial conditions generated using the Zeldovich
Approximation. 
Max. Time-step is the initial global
time-stepping used, which is of order $1\%$
of the Hubble time (i.e, $d \log a=0.01$, being $a$ the scale factor).
The number of global time-steps to complete the runs were $N_{steps}
\simgt 2000$ in all cases, except for the MICE-GC which took 
$\simgt 3000$ time-steps.
Their cosmological parameters were kept constant throughout the runs
(see text for details).}
\label{simtab}
\end{table*}


\section{The MICE Grand Challenge Simulation}   
\label{sec:sim}

\begin{figure}
\begin{center}
\includegraphics[width=0.5\textwidth]{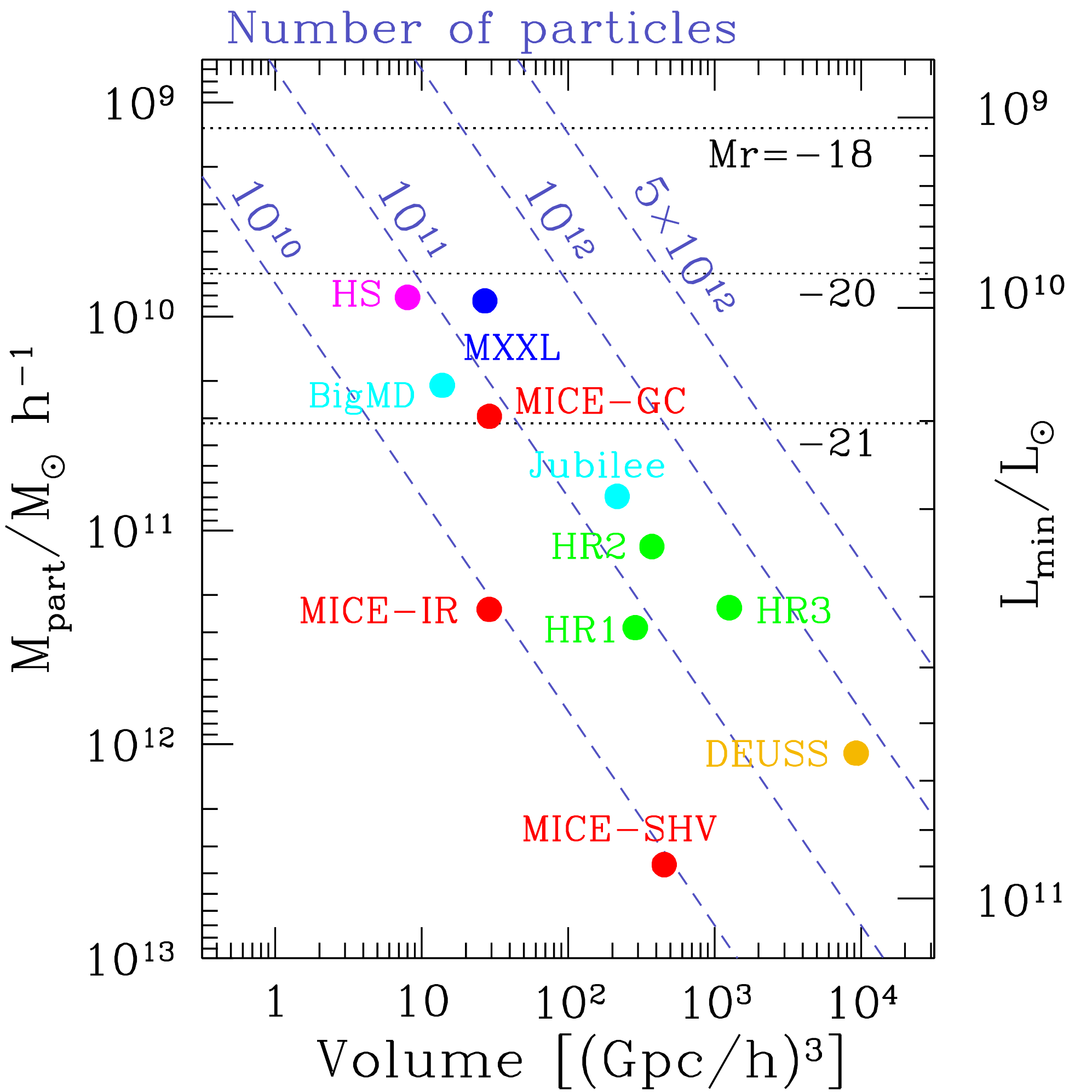} 
\caption{State-of-the art in cosmological simulations: performance in
 terms of survey volume and particle mass, or equivalently, faintest
 $L_{min}$ galaxy luminosity (or absolute magnitude in r-band) reached
according to an HOD galaxy assignment scheme to populate halos with
$100$ or more dark-matter particles.}
\label{fig:cosmosims}
\end{center}
\end{figure}

We have developed a set of large N-body simulations with Gadget-2 \citep{springel05}
on the Marenostrum supercomputer at BSC \footnote{Barcelona Supercomputing Center, www.bsc.es}.
We shall name them MICE (MareNostrum - Instituto de Ciencias del Espacio) 
simulations hereafter. Further details about the MICE simulations can
be found in the project website, \texttt{ www.ice.cat/mice}. 

In this paper we shall focus on the MICE {\it Grand Challenge}
simulation (MICE-GC hereafter). This simulation
contains $4096^3$ dark-matter particles in a box-size of $L_{box}=
3072 \Mpc$, and assumes a flat concordance LCDM model with $\Omega_m=0.25$,
$\Omega_\Lambda =0.75$, $\Omega_b=0.044$, $n_s=0.95$, $\sigma_8=0.8$
and $h=0.7$, consistent with the best-fit cosmology to WMAP 5-year
data \cite{2009ApJS..180..306D}. 
The resulting particle mass is $m_p = 2.93 \times 10^{10} \Msun$ and  
the softening length used is, $l_{soft} =  50 \rm{kpc/h}$. 
We start our run at $z_i = 100$ displacing particles using the
Zeldovich dynamics.
The initial particle load uses a cubic mesh with $4096^3$ nodes.
The entry Max. Time-step in Table~1 is the initial global
time-stepping used, which is of order $1\%$
of the Hubble time (i.e, $d \log a=0.01$, being $a$ the scale factor).
The number of global time-steps to complete the run was $N_{steps}
\simgt 3000$, and we write the lightcone on-the-fly in 265 steps
  from z=1.4 to z=0.

In Table \ref{simtab} we describe the Gadget-2 code parameters used in the MICE
simulations discussed in this paper: the MICE Grand-Challenge (MICE-GC),
the Intermediate Resolution (MICE-IR; \cite{fosalba08}) and the Super-Hubble
Volume (MICE-SHV; \cite{crocce10}).



\begin{figure*}
\begin{center}
\includegraphics[width=0.33\textwidth]{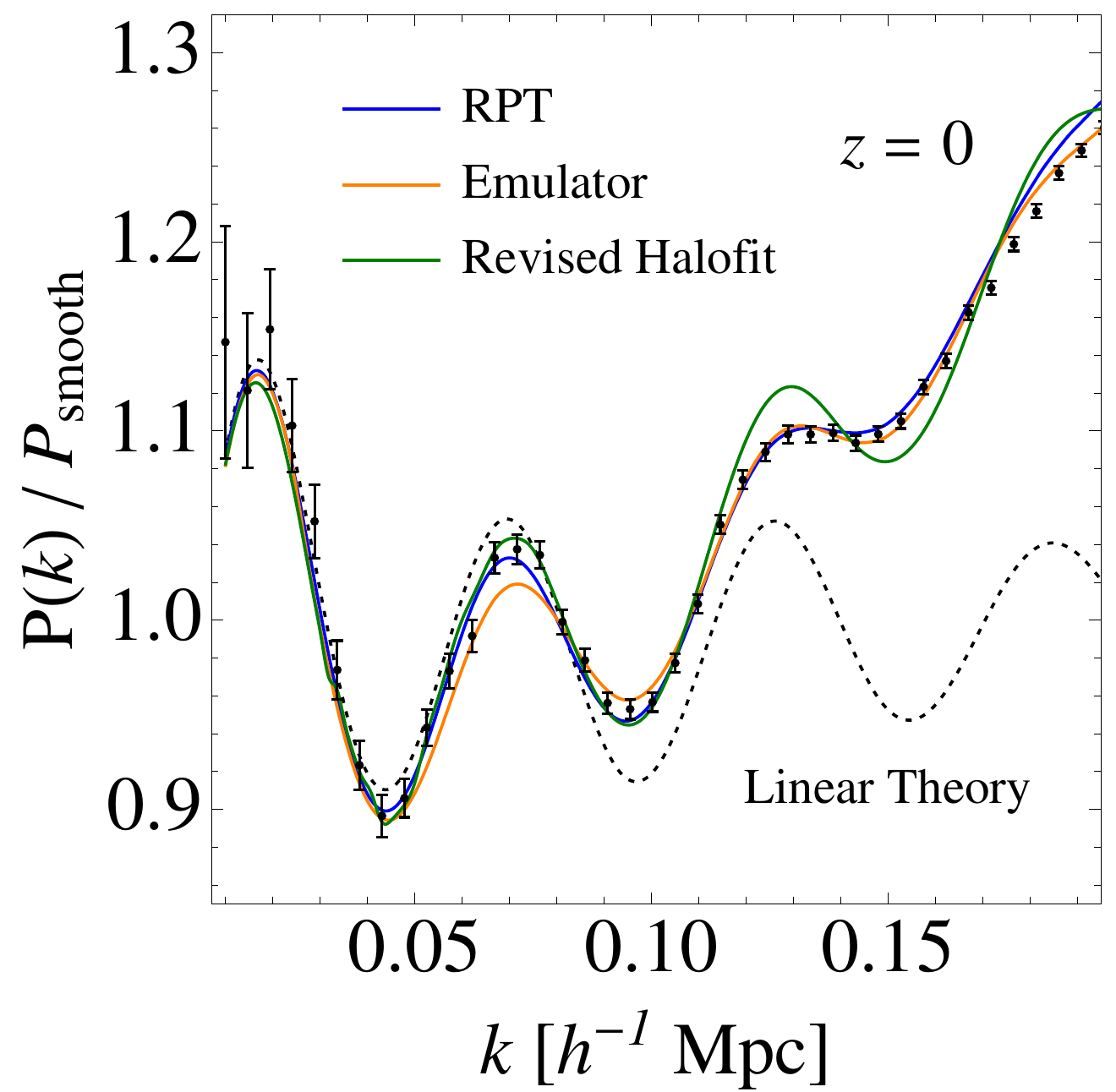} 
\includegraphics[width=0.33\textwidth]{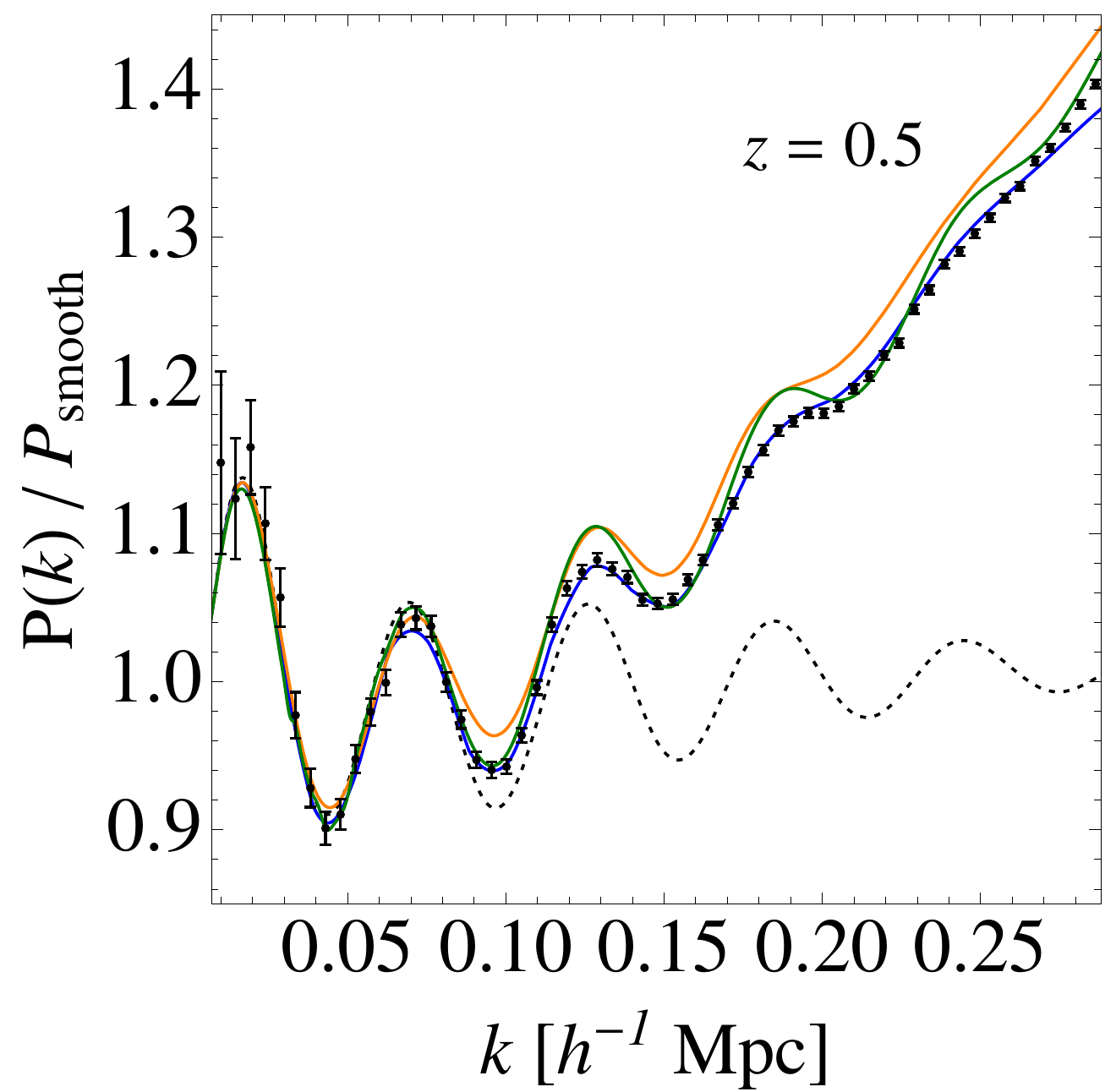} 
\includegraphics[width=0.33\textwidth]{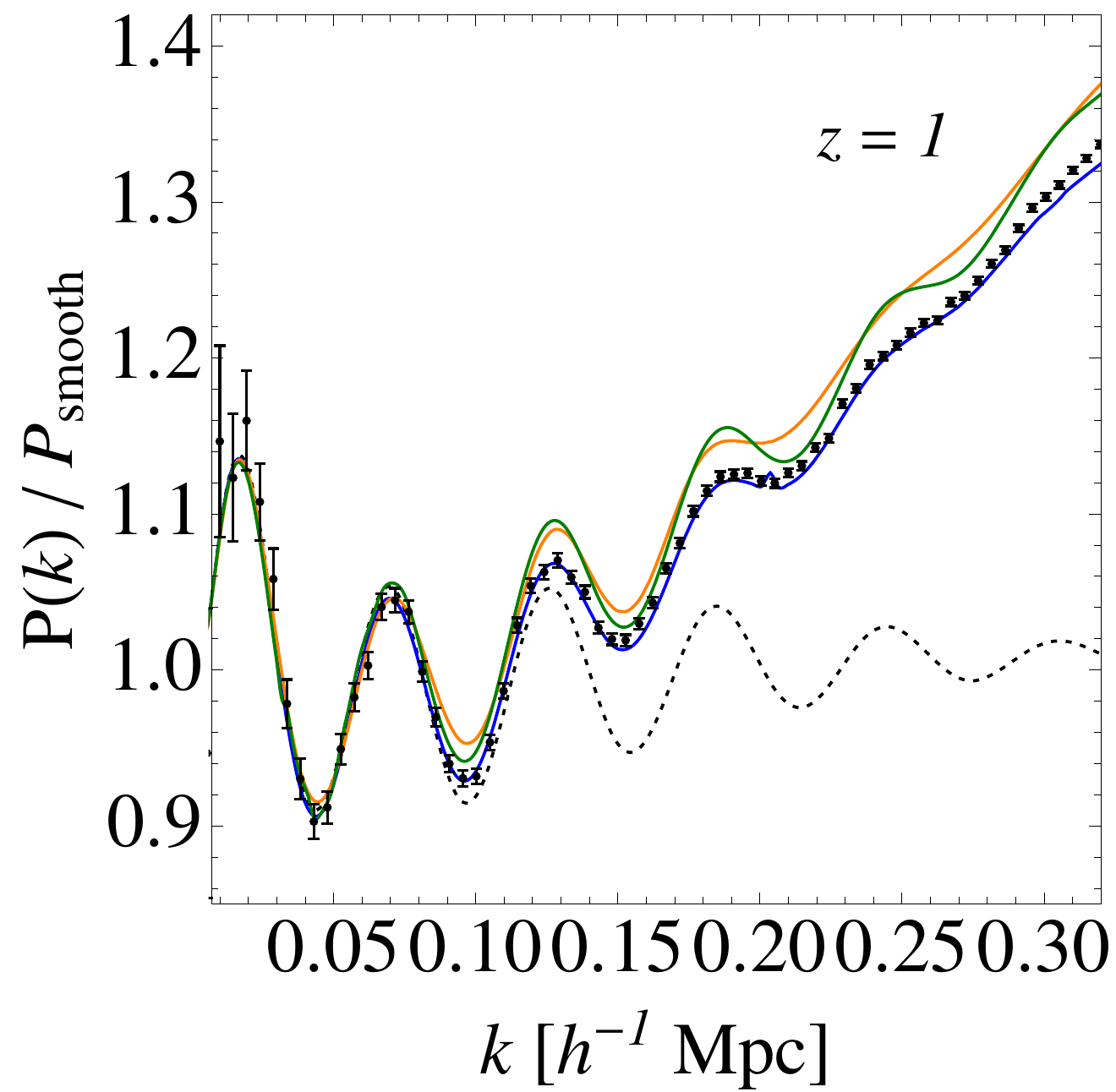} 
\caption{Baryon Acoustic Oscillations (BAO) measured in MICE-GC  (black
  symbols with error-bars) power spectrum compared to 
  the theory prediction from Renormalized Perturbation Theory, RPT 
(blue line, see Crocce and Scoccimarro 2008) and the
latest numerical fit from the Coyote Emulator (orange line, Heitmann
et al. 2013) and the revised Halofit (green line, Takahashi et al. 2012). 
The RPT model at two loops reproduces very well the BAO in the simulation across
redshifts (each panel is shown up to the maximum $k$ where RPT is valid). 
In turn at $z=0$ the Emulator yields a very good match with MICE-GC 
(except for the amplitude of the first peak with a difference
$\lesssim 2\%$). At $z=0.5,1$ the broad-band power has the correct
shape but is $2\%$ (systematically) 
above the N-body. The revised halofit also agrees with MICE-GC
at the $2\%$ level at these redshifts but the amplitude of the
oscillations are somewhat too large. Displayed error-bars assume
Gaussian fluctuations, $\sigma_P = \sqrt{2/n_{\rm modes}} P_k$, 
but we take $P_k$ to be the non-linear spectrum (see text).} 
\label{fig:BAO}
\end{center}
\end{figure*}

We construct a lightcone simulation by replicating the simulation
box (and translating it) around the observer, thus
allowing to build all-sky lensing outputs with negligible repetition up to $z_{max}=1.4$.
The methodology used to
develop all-sky (spherical) lightcone outputs is described
in \citep{fosalba08}, where further details are given on how the
all-sky lensing maps in the Born approximation are constructed using
pixelized 2D projected dark-matter density maps.
Figure~\ref{fig:micesphere} shows an image of the all-sky lightcone at
$z=0.6$. The nested zooms illustrate the wide dynamic range sampled
by the simulation. In particular, thanks to its combination of large
volume and good mass resolution, the MICE-GC simulation is able to 
sample from the largest (linear) scales encompassing nearly
all the observable universe down to  fairly small (non-linear)
scales with few percent accuracy, as we shall show in this work.

A particularly challenging aspect of any large-volume cosmological
simulation is the ability to accurately sample small-scale
(non-linear) power while still sampling a large volume. 
This ability is determined by the particle mass
(or density of particles) in the simulated volume. 
In what follows, we shall refer to
the impact of the particle mass on clustering observables as 
{\it mass-resolution effects}.

In this paper we introduce this new very large cosmological simulation as
a powerful tool to model accurately current and upcoming deep wide-area
astronomical surveys such as 
DES\footnote{\texttt{www.darkenergysurvey.org}}, 
HSC\footnote{\texttt{www.naoj.org/Projects/HSC}}, 
Euclid\footnote{\texttt{www.euclid-ec.org}}, 
DESI\footnote{\texttt{desi.lbl.gov}},
HETDEX\footnote{\texttt{hetdex.org}}, 
LSST\footnote{\texttt{www.lsst.org}}, 
WFIRST\footnote{\texttt{wfirst.gsfc.nasa.gov}}, among others.

We test the ability of MICE-GC to model these large-surveys
on the smaller scales and to what extent it resolves the small-mass halos
inhabited by the faintest galaxies these surveys will observe.
Figure~\ref{fig:cosmosims} shows how MICE-GC compares to the largest
simulations currently available in performance to sample large
cosmological volumes and capture, at the same time, low enough
luminosity galaxies $L_{min}$, or equivalently, large enough r-band absolute
magnitude $M_r$. The relation between minimum halo mass and minimum galaxy
luminosities modeled, as shown in the Figure, assumes a sub-halo abundance matching galaxy assignment scheme on
well-resolved dark-matter halos containing at least 100 particles.
We show the following simulations: Millennium XXL (MXXL; \cite{angulo12}), 
Horizon Runs (HR; \cite{kim09,kim11}), Horizon Simulation
(HS; \cite{teyssier09}), DEUSS \citep{alimi12},  Jubilee
\citep{watson13}, BigMultiDark\citep{BigMD},
MICE Intermediate Resolution (MICE-IR; \cite{fosalba08}), MICE Super-Hubble-Volume
(MICE-SHV; \cite{crocce10}) and the MICE-GC simulation. 

This suite of simulations includes numbers of particles that span from about 10
billion up to 1 trillion particles, already accesible in the largest
supercomputers around the world. This figure shows the trade-off
between high mass resolution and large volume sampling what tends to
distribute the most competitive simulations to date along the dashed
lines shown, depending on the number of particles used.
As shown in Fig. \ref{fig:cosmosims}, the MXXL is the
state-of-the art cosmological simulation in terms of sampling as large
a volume as possible with a high mass resolution. Our simulation
compares well to it, as it has a comparable volume (3.072 Gpc/h) and only 4 times
lower mass resolution ($2.93\times 10^{10} \Msun$).

As an example, in order to resolve $M_r=-18$ galaxies
one would need to develop a simulation with a (4 Gpc/h)$^3$ box-size that
includes about 5 trillion particles (i.e, $16384^3$). This is one order
of magnitude larger than the MXXL and almost two orders of magnitude
bigger than e.g, the MICE-GC, which are among the largest
simulations completed to date.

\section{3D Clustering}
\label{sec:3dclustering}

\begin{figure}
\begin{center}
\includegraphics[trim= 2cm 0.4cm 1.0cm 0cm, clip=true, width=0.5\textwidth]{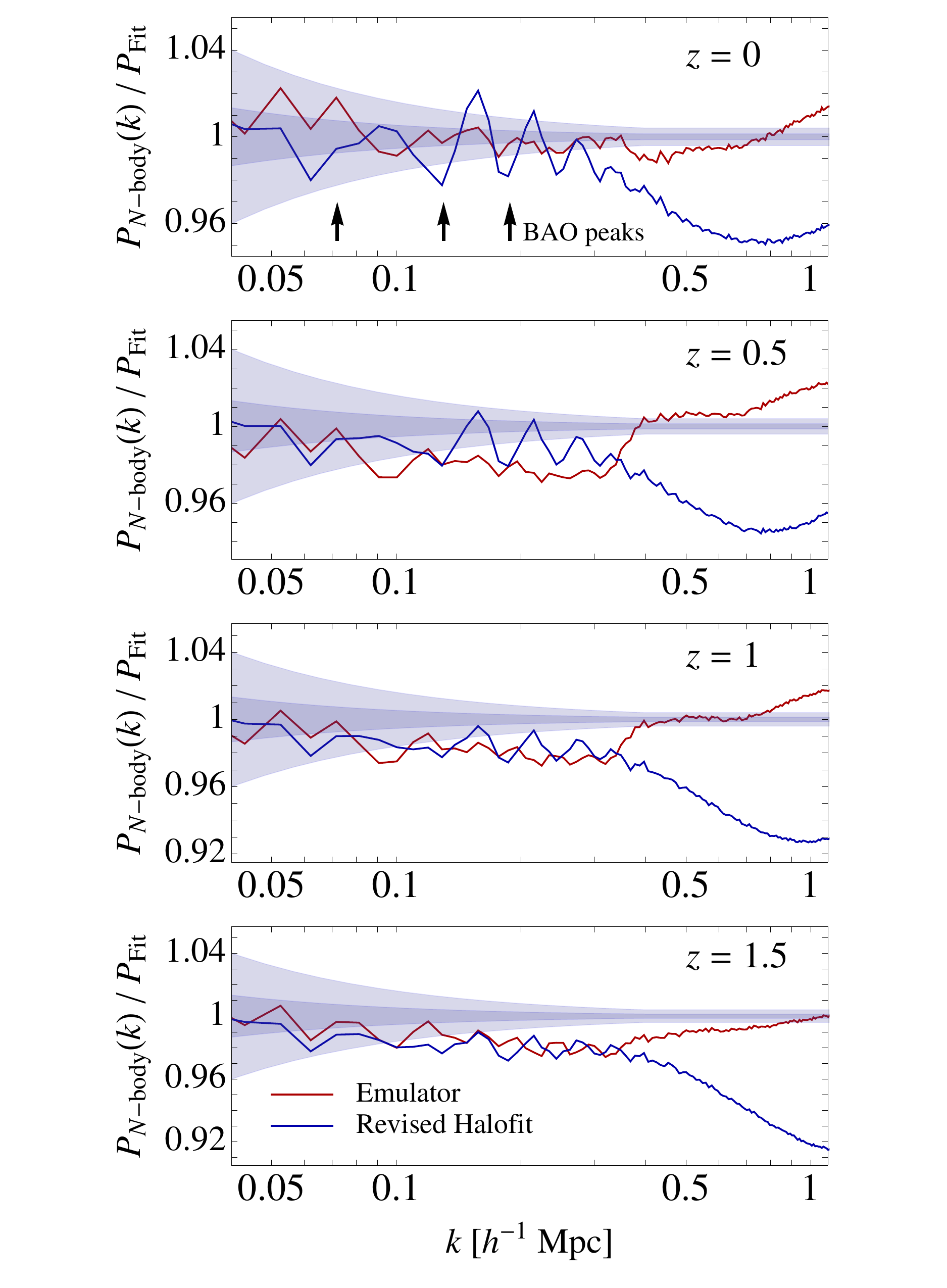} 
\caption{Matter power spectrum in MICE-GC at
  several comoving outputs compared to the latest available numerical
  fits, the
  revised Halofit (Takahashi et al. 2012) and the Coyote Emulator
  (Heitmann et al. 2013). The shaded regions show the $1-\sigma$ and
  $3-\sigma$ error bars 
  given the box-size and the binning (see text for details). The overall matching with the emulator
  is at the level of $2\%$ (but systematically below), with a  jump of
  $\lesssim 3\%$ at $k\sim 0.3-0.4\,{\it h}^{-1}{\rm Mpc}$. In turn, the revised Halofit
  over-predicts the power on small scales by $5\%$ to $8\%$ (see text for a more
  detailed discussion). Vertical arrows show BAO peaks for reference.} 
\label{fig:Coyote}
\end{center}
\end{figure}

\subsection{Power spectrum}
\label{sec:powerspectrum}

One of our main goals is to study the large scale clustering
with high precision, in particular the baryon acoustic oscillations (BAO). Hence Fig.~\ref{fig:BAO}
shows the matter 3D power spectrum measured in MICE-GC 
at large (BAO) scales for three comoving outputs, $z=0,0.5$ and $1$ 
(divided by a smooth broad-band power). For comparison we included
linear theory, the Renormalized Perturbation Theory
prediction as presented in \cite{RPT}, and the
numerical fits from \cite{heitmann13} (\ie the Coyote Emulator) and
\cite{takahashi12}, which we shall name the {\it revised halofit}.
The prediction from RPT at two-loops
reproduces very well the region of BAO, thus cross validating the modeling
and the N-body precision in describing this feature. We find the Coyote Emulator to give a good
description of the overall BAO features in our run, but with a systematic $2\%$
excess (broad-band) amplitude at $z=0.5,1$. The revised Halofit yields a similar
level of agreement but in that case the amplitude of the oscillatory
features appear too large.

In Fig.~\ref{fig:BAO} displayed error bars corresponds to the FKP formula \citep{FKP}:
\begin{equation}
\sigma_P = \sqrt{\frac{2}{n_{\rm modes}}} P
\label{eq:FKP}
\end{equation}
but using $P$ as the non-linear power spectrum (and no shot-noise, as
this is negligible for MICE-GC). In Eq.~\ref{eq:FKP} $n_{\rm
  modes}= 4 \pi \, \delta k \, k_f$ is the number of modes, $k_f= 2\pi /
L_{box}$ the fundamental mode for the given boxsize, and $\delta k$
the bin width in $k$-space. 
We have tested that this yields a
good estimate for the statistical errors at BAO scales with an independent ensemble
of simulations \citep{RPT}. Similar conclusions are
found in \cite{2008MNRAS.383..755A}. At $k \ge 0.4 \Mpc$ we find that the relative
error $\sigma_P/P$ flattens out to a roughly constant value (see also \cite{2009ApJ...700..479T,2009MNRAS.400..851S}). 

In turn, Fig.~\ref{fig:Coyote} shows the MICE-GC 2-pt clustering
measurements in the transition to the highly 
nonlinear regime and how they compare to the revised Halofit and the
Coyote Emulator on those scales 
\footnote{We restrict the comparison to scales ($k \sim 1 \kvecMpc$)
where shot-noise in MICE-GC is negligible. In addition the 
power spectrum estimation was done using a CIC assignment scheme 
with a large mesh ($k_{\rm Ny} = 2\kvecMpc $) which, after correction
from the FT of the assignment function, is sufficient to avoid 
aliasing effects on the scales of interest.}. Different panels correspond to different comoving outputs,
namely $z=0,0.5,1$ and $1.5$ (top to bottom). As discussed above the
Emulator is systematically above our measurement on scales $k \lesssim
0.3 \kvecMpc$ by about $2\%$ (for $z>0$). Towards smaller scales we
find a pronounced jump in the prediction at intermediate redshifts and
$k \sim 0.3-0.4 \kvecMpc$ of $\sim 3\%$. The overall matching with
MICE-GC measurements stays at the level of $\pm 2\%$ up to $k \sim
1\kvecMpc$. The revised Halofit is slightly above our measurements
across scales, but particularly at the ``one-halo'' regime beyond $k
\sim 0.3 \kvecMpc$, where differences rise to about $5-8\%$ depending
on redshift. In
Fig.~\ref{fig:Coyote}, and following the discussion after Eq.~(\ref{eq:FKP}), we
display $1$ and $3\,\sigma$ error bars with shaded regions using Eq.~(\ref{eq:FKP}) for $k \le 0.4\kvecMpc$ and its
value $(\sigma_P/P)|_{k=0.4}$ for $k \ge 0.4\kvecMpc$.

We should note that on small scales ($k \sim 1\kvecMpc$) our run might
suffer from percent level systematics effects that could impact the discussion above.
For example setting the size of the Particle-Mesh grid (the PMGrid
parameter) equal to the number of particles (as done in MICE-GC, see
Table \ref{simtab}), instead of larger, can yield excess power on
quasilinear scales of $\lesssim 1\%$ \citep{smith12}. 
In turn, transients from initial conditions suppress power on these scales. For MICE-GC, with $z_i=100$, we expect this to
be a small effect nonetheless.  For an order of magnitude we have
estimated the relative transient effect
(i.e. $P(k,z_i=100)/P(k,z_i=\infty)$) using perturbation theory (PT,
see Fig. 6 in \cite{2LPT}) and found a $\sim (1-1.5) \%$ effect at scales $k \sim 0.5-1 \kvecMpc$.

\begin{figure}
\begin{center}
\includegraphics[trim = 0cm 2.1cm 0cm 0cm,width=0.42\textwidth]{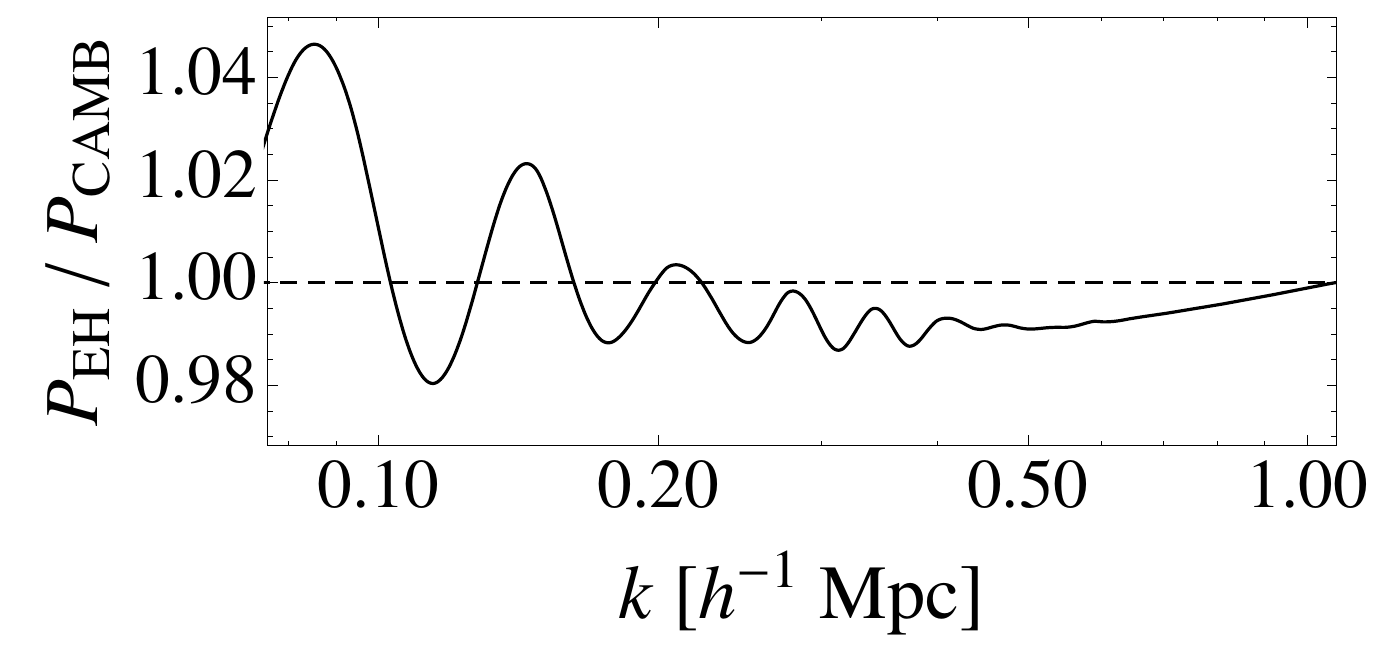} 
\includegraphics[width=0.42\textwidth]{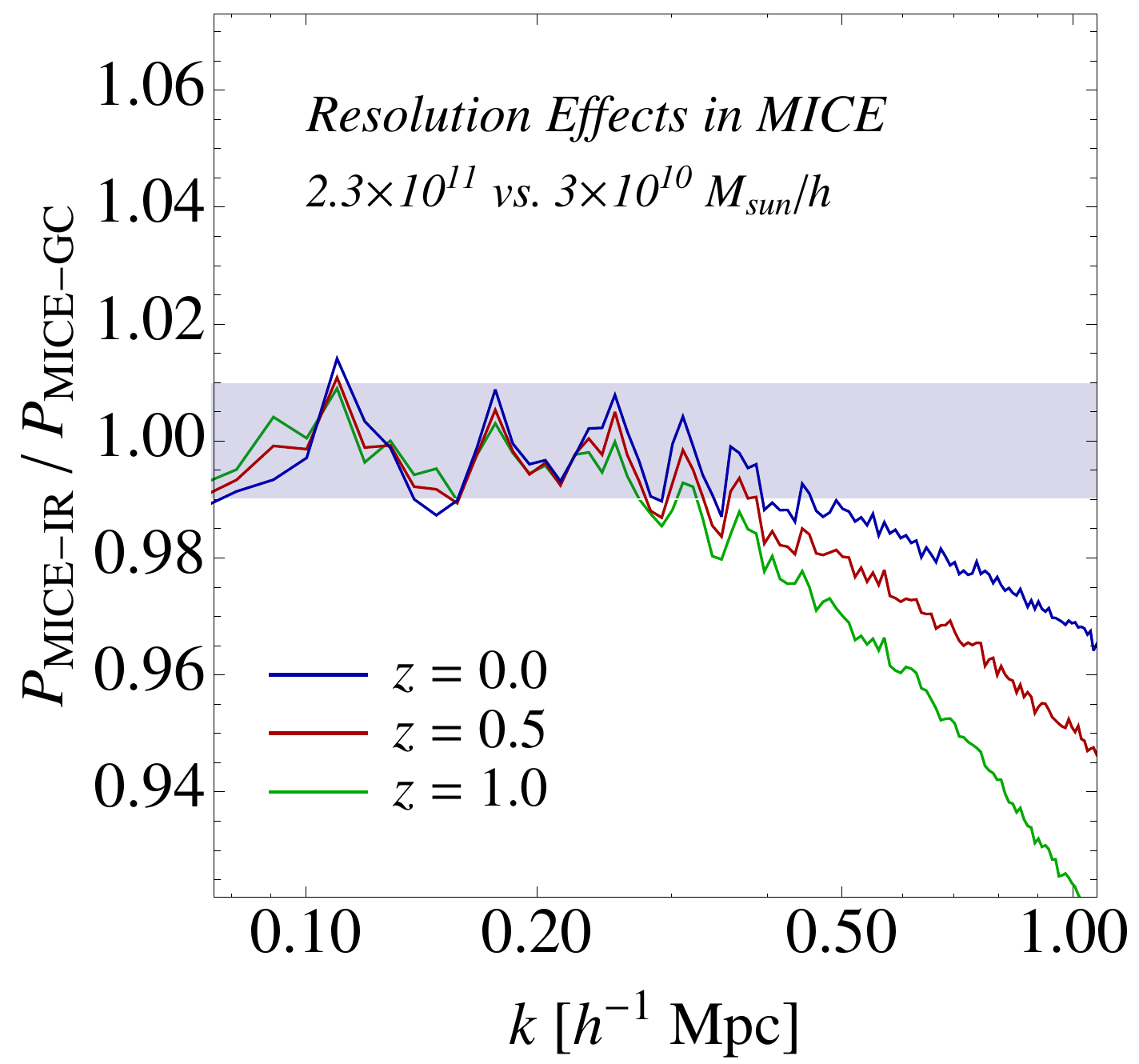} 
\caption{{\sc Top Panel: }Ratio of the initial Eisenstein \& Hu linear power spectrum
  used in MICE-IR to the one used in MICE-GC from {\tt CAMB}, both for the
  same (MICE) cosmology. {\sc Bottom Panel: } suppression of nonlinear
  structure formation due to particle mass resolution seen through the 
  ratio of the power spectrum measured in MICE-IR ($m_p \sim 2.3\times 10^{11}
  \Msun$) to the one in MICE-GC ($m_p
  \sim 3\times10^{10}\Msun$) at $z=0,0.5$ and $1$. MICE-IR measurements were corrected assuming a Poisson
  shot-noise and the slight difference in initial spectra was divided out. If we do
  not correct for shot-noise we find these ratios to be almost
  independent of redshift and to resemble the $z=0$ case shown by the
  blue line.} 
\label{fig:Pkmassresolution}
\end{center}
\end{figure}

We next turn to discuss the impact of particle mass resolution in the
generation of nonlinear structure by comparing the clustering in
MICE-GC to that in MICE-IR, a factor of 8 worse resolution
run. This lower resolution simulation was already presented and
extensively validated in \cite{fosalba08}, and \cite{crocce10}
\footnote{ note that in Crocce et al. 2010, the MICE-IR 
simulation was named MICE3072, see their Table 1 for details}. We note that
both simulations were done with almost the same run
parameters (see Table~\ref{simtab}) with the main difference being the
initial transfer function. MICE-IR
used the Eisenstein $\&$ Hu (EH from now on) approximation \citep{eisenstein98} while MICE-GC used an
(exact) {\tt CAMB} output  \footnote{\texttt{http://camb.info/}}. The differences between these two initial spectra, shown in
the top panel of Fig.~\ref{fig:Pkmassresolution}, are however
very small: $\lesssim 4\%$ at BAO positions and within $2\%$ up to $k\sim 50\kvecMpc$ (although only $k
< 1 \kvecMpc$ is plotted).

The resolution study is given in the bottom panel of
Fig.~\ref{fig:Pkmassresolution}. It shows that the nonlinear power
spectrum measured in MICE-IR is suppressed compared to that
in MICE-GC, by $3\%,5\%$ and $\sim 8\%$ at $k=1\kvecMpc$ and $z=0,0.5$ and $1$ respectively.
These differences are somewhat expected because particle mass resolution
 impacts how well small-scale power is sampled from the simulated dark-matter
distribution. The lower the particle mass, the better small scales
are sampled. In the language of the halo model,
a smaller particle mass is able to resolve smaller-mass halos
that contribute to the power spectrum on correspondingly smaller scales.
Notice that the effect can be significant even on rather large-scales
($1\%$ to $2\%$ at $k \sim 0.4\kvecMpc$). In Fig.~\ref{fig:Pkmassresolution} the MICE-IR power spectra have been corrected
for finite particle number assuming a Poisson shot-noise (while
MICE-GC has negligible noise on these scales). If we have
not done so the $P(k)$ ratio at different redshifts would have resembled the
$z=0$ case in Fig.~\ref{fig:Pkmassresolution}, i.e. a resolution effect being $\sim 2\%-3\%$ at $k=1\kvecMpc$.
One difference between the simulation runs that can impact part of
the effect shown in Fig.~\ref{fig:Pkmassresolution} is the
starting redshift. Because MICE-IR started at a lower redshift than MICE-GC
($z_i=50$ vs. $z_i=100$, respectively) its clustering is expected to be
suppressed at non-linear scales. However we have estimated this to be a $1\%$
effect in the k-range $0.5-1 \kvecMpc$ for the redshifts shown (using PT, see \cite{2LPT}).

\subsection{2-point correlation function}
\label{sec:2pcf}
 
\begin{figure*}
\begin{center}
\includegraphics[width=0.45\textwidth]{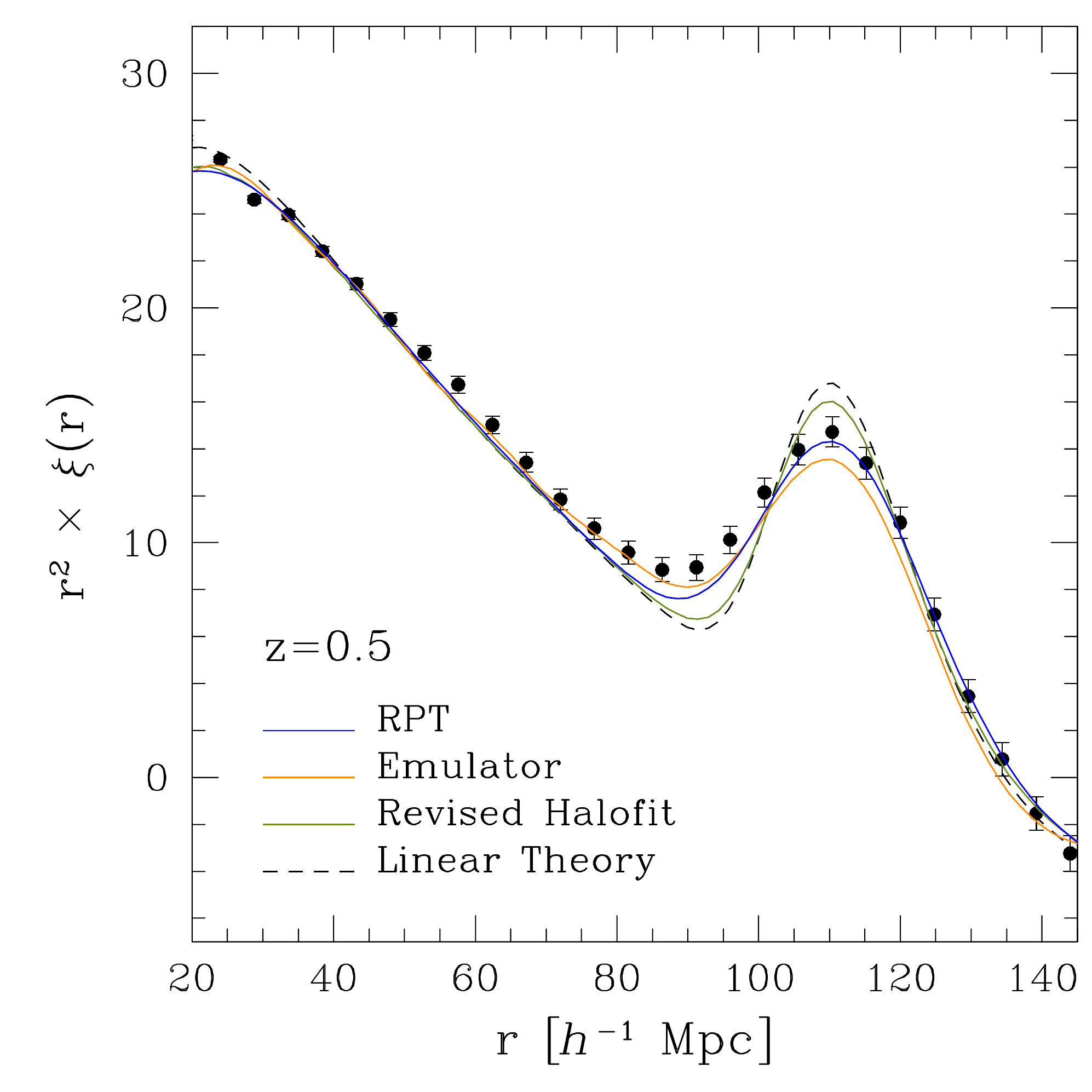} 
\includegraphics[width=0.46\textwidth]{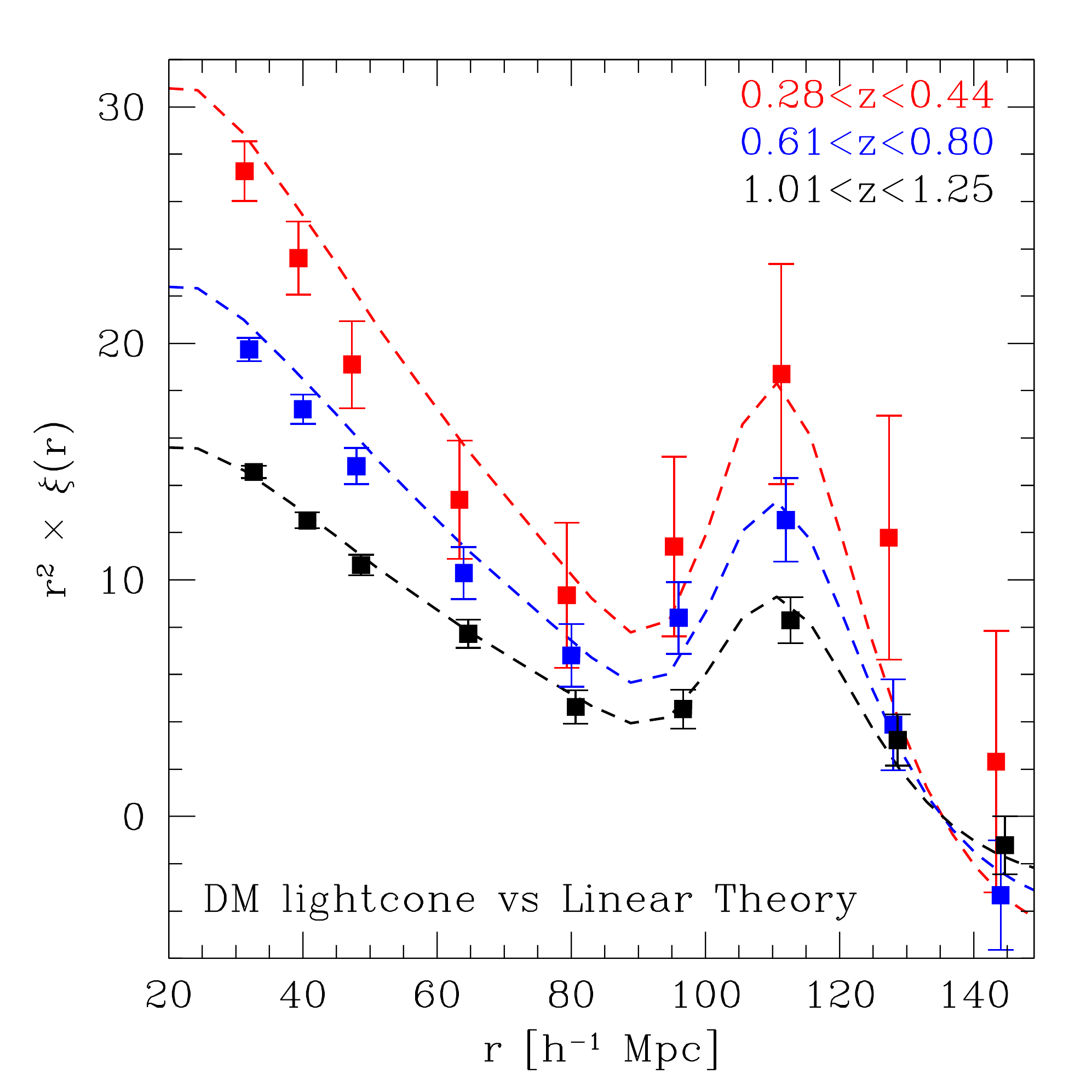}
\caption{{\sc Left Panel:} Dark-matter correlation function in MICE-GC at $z=0.5$ at
  large-scales (the Fourier counterpart to the middle panel of Fig.~\ref{fig:BAO}). We include the
Fourier transform of the models shown in Fig~\ref{fig:BAO}: RPT (blue,
Crocce and Scoccimarro 2008), Emulator
(orange, Heitmann et al. 2013), the revised Halofit (green,
Takahashi et al. 2012) and linear theory (dashed black). The later is the one that shows
the worse agreement with our N-body at BAO scales (in the amplitude of
the feature), while RPT and Emulator are in better agreement. Similar
conclusions are reached at $z=0$ and $1$.
{\sc Right Panel:} Corresponding results for difference redshift slices over
one octant of 
the MICE-GC  lightcone output  (also in real space) smoothed over
$8 \Mpc$ pixels and $16 \Mpc$  bins in pair separation $r$.
Dashed lines are the smoothed linear theory predictions (which resemble
non-linear predictions). Note how the errors are larger (and more realistic)
because of the smaller volume available in the lightcone.}
\label{fig:xi3D}
\end{center}
\end{figure*}

To complement the previous sections we show in left panel of  Fig.~\ref{fig:xi3D}
the spatial correlation function of dark matter particles in the $z=0$
comoving output of MICE-GC. This is basically the Fourier counterpart
of Fig.~\ref{fig:BAO} but it is still interesting to see how any
mismatch between the measurements and the theory in that figure translate to
configuration space. The BAO wiggles from the revised Halofit \citep{takahashi12} which are too pronounced in
Fourier Space also yield the wrong amplitude for the correlation
turn-over at $\sim 90\Mpc$ and the BAO peak at $110\Mpc$. Both RPT
\citep{RPT} and Coyote \citep{heitmann13} seem to agree better with
MICE-GC. All models agree very well with themselves and the
measurements on smaller scales, $20 \Mpc \le r \le
70 \Mpc$. For concreteness we focused on $z=0.5$ but similar
conclusions are reached for $z=0$ and $z=1$.

Right panel of Fig.\ref{fig:xi3D}, shows the correlation function 
measurements for different
redshifts in the lightcone. 
The top-hat filter is a cubical cell of side 8Mpc/h. We split the correlation in  bins
of 16 Mpc/h to reduce the errorbars. So the smoothing is not necesary, but it speeds
up the codes, which are the sames we use for the 3-pt correlation, for which the
execution time reduction is critical. The redshift bin is given in complete cell sizes which also reduces
the impact of the boundaries. The redshifts depth could add some
effects due to selection in redshift space, but we have checked that mean results do not change much (within
errors) for other bin widths.
This smoothing makes the linear and non-linear results look quite closer on BAO scales.
On the largest scales
there is good agreement, within the errors, with the linear theory 
predictions (dashed lines), specially at the larger redshifts. At the lower redshifts, 
non-linear effects are more important (distorting the shape of the BAO
peak and the amplitude around $r \simeq 40-80 \Mpc$) and errors are larger because of 
the smaller volume in the lightcone. Given these errors and the smoothing, it is hard to evaluate if
there are additional lightcone distortions in addition to the
non-linear effects that we find in the comoving outputs. 

\section{Angular Clustering in the lightcone}
\label{sec:cls}

Following the approach presented in \cite{fosalba08}, 
we construct a lightcone simulation by replicating the simulation
box (and translating it) around the observer. 
Given the large box-size used for the MICE-GC simulation,
$L_{box}=3072 \Mpc$,  this approach allows us to build
all-sky lightcone outputs without significant repetition up to $z_{max}=1.4$.
Then we decompose the lightcone volume,  in the range $0 < z <1.4$, into a set of 265 all-sky
concentric spherical (or radial) shells around the observer,  with a constant width of
$\sim 70$ Mega-years in lookback time. This width corresponds to about
8 Mpc/h at $z\simeq 0$ and 15 Mpc/h at $z\simeq 1.4$.
Lightcone outputs are written
on the fly for each of these sphericall shells. We linearly
interpolate particle positions using their velocities at the lightcone timestep or radial shell.

\subsection{Angular Clustering in Real Space}
\label{sec:clreal}

\begin{figure*}
\begin{center}
\includegraphics[width=0.95\textwidth]{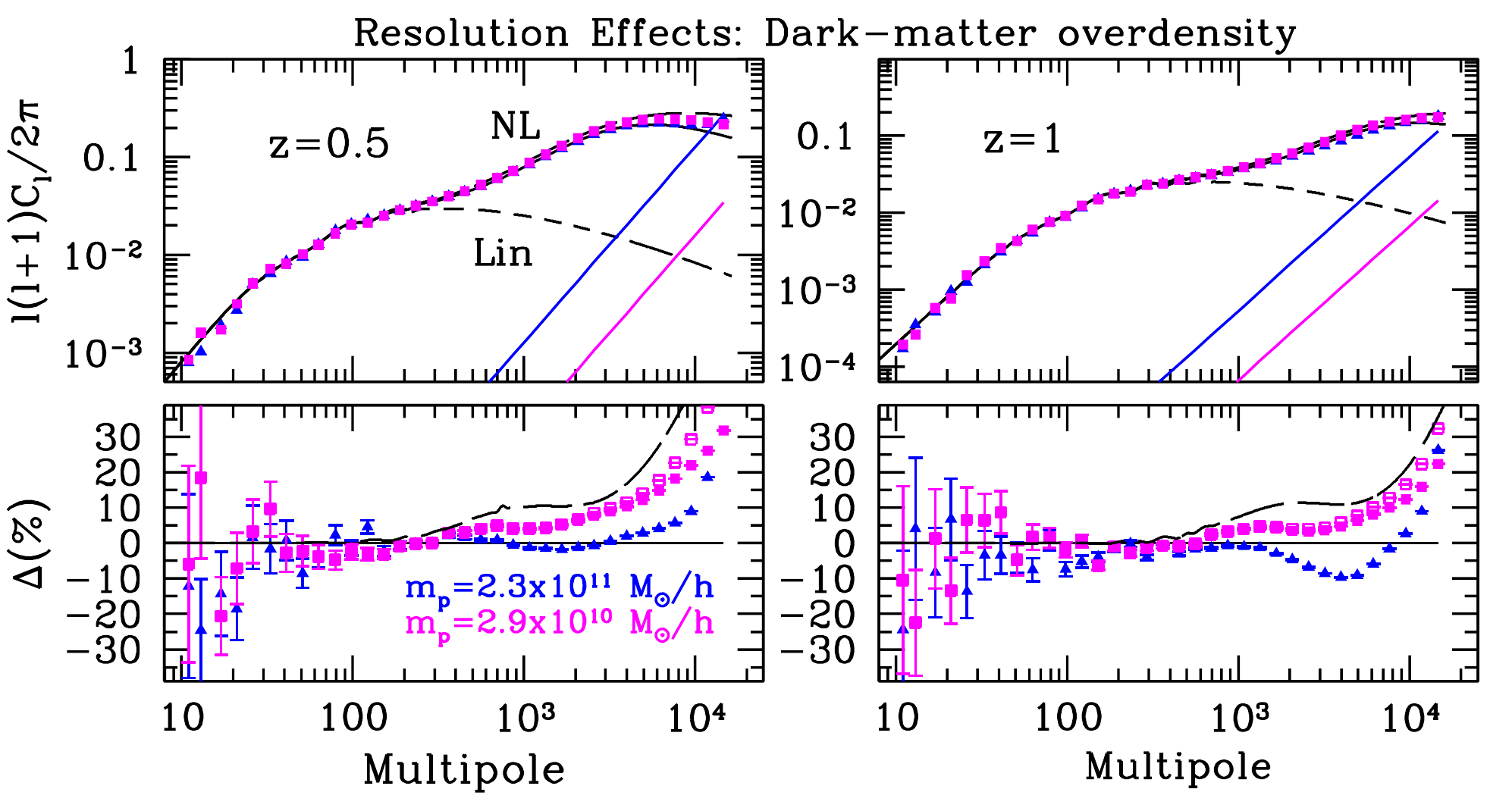} 
\caption{Angular power spectrum of the projected dark-matter
overdensity field in redshift (all-sky lightcone) slices. Left panels show power spectra
for a broad z-bin slice at $z=0.5$ and $\Delta z=0.1$, whereas the right panels show the case
for $z=1$ with the same binwidth. 
Dashed and solid lines show linear theory and non-linear
fit (Halofit) predictions, respectively. 
Long dashed lines shows the {\it revised Halofit} 
\citep{takahashi12}
which shows a clear excess of power with respect to the {\it old Halofit}
\citep{smith03}. 
Symbols show measurements from
simulations and the solid lines at the bottom right display the
estimated shot-noise levels.
Lower panels show relative deviations with respect to the the old Halofit prediction. 
For clarity, in these lower panels, we only
show power spectra both with (open symbols) and without (filled
symbols) shot-noise for the MICE-GC simulation, where
shot-noise does not affect small-scale power significantly. Mass resolution effects
at the $20-30\%$ level are observed for multipoles $\ell > 10^3$.}
\label{fig:clmassres}
\end{center}
\end{figure*}

\begin{figure*}
\begin{center}
\includegraphics[width=0.95\textwidth]{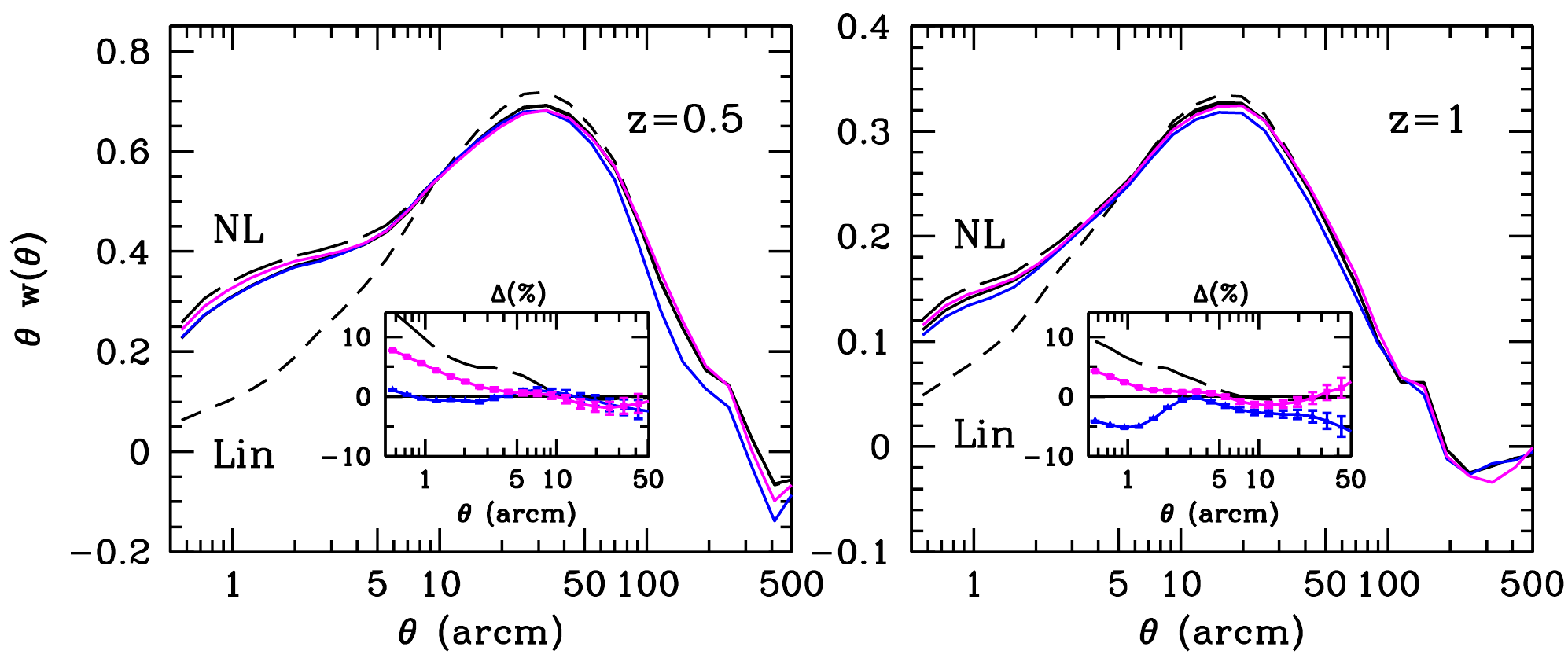} 
\caption{
Same as Fig.~\ref{fig:clmassres} but for the angular 2-point
correlation function. In order to better compare to theory
predictions we use
lines (instead of symbols) to display MICE-GC measurements. The inset
panels display the ratio of
measurements (and revised halofit) to the Smith et al. (2003) halofit
prediction. We zoom into smaller angular scales to highlight mass resolution
effects.}
\label{fig:wmassres}
\end{center}
\end{figure*}


Next we validate our simulation dark-matter outputs in 2D. In order to do this 
we use the all-sky 3D lightcone dark-matter density, decomposed into
narrow spherical redshift slices, and projected onto 2D pixelized maps
using the Healpix tessellation\footnote{{\texttt
    http://sourceforge.net/projects/healpix}}\citep{gorski05}.
As a first statistical test, we compare the 2-point statistics
to theoretical predictions.
In general, one defines the angular power spectra of scalar fields on
the sphere, $C_{\ell}$, as follows,
\beq
<X^{\ast}_{\ell m} Y_{\ell^{\prime}m^{\prime}}> =  \delta_{\ell \ell^{\prime}}\delta_{m m^{\prime}} \,C_{\ell}^{XY} \, ,
\eeq
where $X$ and $Y$ are two arbitrary 2D fluctuating fields.

Figure ~\ref{fig:clmassres} shows the angular power spectrum for
all-sky projected dark-matter overdensity in narrow redshift slices, for
the MICE-GC, compared to a previous lower-mass resolution run, the
MICE-IR. 
Particle shot-noise, shown as solid lines at the right of
the upper panels, has been subtracted from the measured power spectra
and displayed theoretical error bars correspond to an (ideal) all-sky survey.
Throughout this paper, we shall use Gaussian theoretical error bars for
the different observables of 2D clustering,
unless otherwise stated, following \cite{cabre07,crocce11}.
We shall stress that this is an optimistic error estimate on small scales, since a significant deviation from the Gaussian
approximation is expected to arise as a result of non-linear
gravitational growth \citep{scoccimarro99}.

As shown in the lower panels of Fig.~\ref{fig:clmassres}, 
using a particle mass 8 times lower,
produces a drop in the measured power at the 
$5\%$ level for 
multipoles $\ell \sim 10^3$, which corresponds to wavenumbers $k \sim
0.75$ at $z=0.5$ and $k\sim 0.5$ for $z=1$ in the flat-sky (or
Limber) limit. The impact of such mass-resolution effect is even larger for higher redshift.
This is consistent with the effect seen in the 3D power
spectrum (see Figure~\ref{fig:Pkmassresolution}). 
The fact that the magnitude of mass-resoution effects (and its
  dependence with redshift) at these quasi-linear scales  are consistent in 3D comoving versus 2D
  lightcone outputs implies that possible lightcone artifacts (e.g,
  interpolation issues between lightcone time-steps) do not affect
  our results.
For even smaller scales
($10^3 < \ell < 10^4$) the difference rises up to $10\%-20\%$,
although here the amplitude of the effect and its scale-dependence start to 
be sensitive to the details of the shot-noise correction.
Besides, at these highly non-linear scales, we cannot exclude possible
  lightcone interpolation artifacts, but again these would be
  difficult to separate from other sources of error such as inaccurate
  shot-noise correction.
Comparing to state-of-the-art numerical fits, the MICE-GC measurements
 in the range $10^3<\ell<10^4$ show an $5-10\%$ excess relative to
 Halofit \citep{smith03}, 
and a $5-10\%$ deficit with respect to the
revised Halofit by \cite{takahashi12}, based on small-box
high-resolution Nbody simulations. This trend of increasing power excess
as one uses higher-resolution simulations indicates that what we
observe on these scales is also consistent with the impact of
mass-resolution on the clustering.

Any inconsistency in the parameters used for the simulations compared above
could question the robustness of our conclusions.
In particular, MICE-GC which uses the exact transfer function
(computed with {\tt CAMB)} to generate the initial conditions, whereas 
MICE-IR employs the Eisenstein $\&$ Hu (EH) approximation to the exact
transfer function. 
However, differences between the exact
and EH power spectrum are typically within $2\%$ for $k\simlt
10$, which corresponds to multipoles $\ell\simlt 10^4$ for $z>0.5$. 
Therefore the observed discrepancy in the angular power spectra cannot be
attributed to this slight inconsistency in the initial conditions.
Other effects on this $\ell$ range, such as transients from initial
conditions (see \cite{2LPT}), are expected to have a negligible impact
for the MICE-GC outputs on these scales,  but a detailed discussion will be
presented elsewhere.

In Figure~\ref{fig:wmassres} we show the equivalent of Figure~\ref{fig:clmassres}
but for the angular 2-point correlation function (A2PCF), \ie the Legendre Transform of the
angular power spectrum,
\beq
w(\theta) = \sum_{\ell} \frac{2\ell+1}{4\pi} C_{\ell}
L_{\ell}(\cos\theta)  ,
\eeq
where $L_{\ell}$ are the Legendre polynomials of order $\ell$. 
It is clearly observed that there
is a transfer of power from large (linear) to small
(non-linear) angular scales as a result of gravitational growth. 
Theory fits predict that the amplitude of this transfer of power, as well as the
angular scale that separates linear and non-linear regime 
increases with decreasing redshift.
In particular, this transition angle is expected to be at $\theta
\simeq 10$ arcminutes for $z=0.5$ and $\theta \simeq 5$
arc-minutes for $z=1$, in good agreement with measurements from simulations.
Note that these scales do not correspond to what is naively expected by
using the relation between multipoles and angular scales in the sky,
$\theta = 180^{\circ}/\ell$. In practice this is because there are
cancellations of nonlinear effects in configuration space which extend
the apparent validity of linear theory to smaller scales \citep{RPT}. Because the
transformation to Fourier (or multipole) space mixes the scales in the
A2PCF for which the cancellation occurs nonlinear effects appear at
smaller $\ell$ values in $C_\ell$. 
This makes the 1-to-1 relation above mentioned not to strictly hold.

In a similar manner mass resolution effects are found to be smaller by roughly
a factor of $2$ in the A2PCF with respect to the angular power spectra
for the same redshift bins and corresponding angular scales if the
1-to-1 relation would hold (see Figures~\ref{fig:clmassres} \& ~\ref{fig:wmassres}).
We find a $5-10\%$ and $10-20\%$ effect for $w(\theta)$ and the
$\Cl$'s respectively for broad redshifts bins in the range
$z=0.5-1$.
On the other hand on linear scales, $\theta
\simgt 30-60$ arcmin depending on redshift,  MICE-IR results
clearly deviate from those of MICE-GC (and theory predictions). This
can be due to a good extent to the
different transfer function used to set up the initial conditions
in both simulations. As shown in Figure~\ref{fig:Pkmassresolution},
there are $\simgt 4 \%$ differences between the initial power spectra of these two
MICE simulations at $k \simlt 0.1 \kvecMpc$, what corresponds to
angular scales of $\theta  = 180^{\circ}/(k r(z)) \simeq 80$ and $45$
arcmin, in the Limber or small-angle limit, for $z=0.5$ and $z=1$,
respectively.

\subsection{Angular Clustering in Redshift Space}
\label{sec:clrsd}

In this section we discuss the impact of peculiar velocities (or
redshift space distortions) in the
angular power spectra and correlation function of dark-matter over-densities.
Redshift space distortions (RSD) do not change the
angular position of a {\it single} galaxy,  nor they change significantly its
measured redshift in photometric surveys. 
However the coherent large-scale motions present in RSD
can have a large impact on the net angular correlation in a {\it sample} of galaxies.
In particular,  when the boundaries of the galaxy sample are defined in redshift space 
(as it happens in a real survey) then large-scale motions move structures across the boundaries in a spatially coherent way. 
In particular, the so-called Kaiser effect \citep{RSD}  produces a
large enhancement of the resulting angular correlations 
(see e.g., \cite{crocce10,nock10}).
Although the use of photometric redshifts tends to counteract the
correlations enhancement due to RSD by smoothing the 2PCF, 
the impact of RSD is still measurable in photometric surveys. 
This impact is seen as a net enhancement of the correlations measured in
photometric surveys with respect to that expected in real space, 
and depends on the particular choice of redshift bin-width, being
larger for narrower bins (as shown in our Figs 9 \& 10 below, or Fig.5 in Crocce et al. 2011).

Below we shall model RSD for both broad and narrow redshift bin widths. 
The former is important for research that aims at measuring the growth
of structure through RSD in photometric surveys
\citep{2007MNRAS.378..852P,2011MNRAS.415.2193R,2011MNRAS.417.2577C,2013arXiv1305.0934A}. 
The later scenario is very relevant for recent studies that 
combine RSD in spectroscopic surveys with weak gravitational
lensing, to break degeneracies between bias and dark energy parameters
\citep{2012MNRAS.422.2904G,2013arXiv1307.8062K,2013arXiv1308.6070D,2013arXiv1307.8062K,2013arXiv1308.4164F}.

\begin{figure*}
\begin{center}
\includegraphics[width=0.49\textwidth]{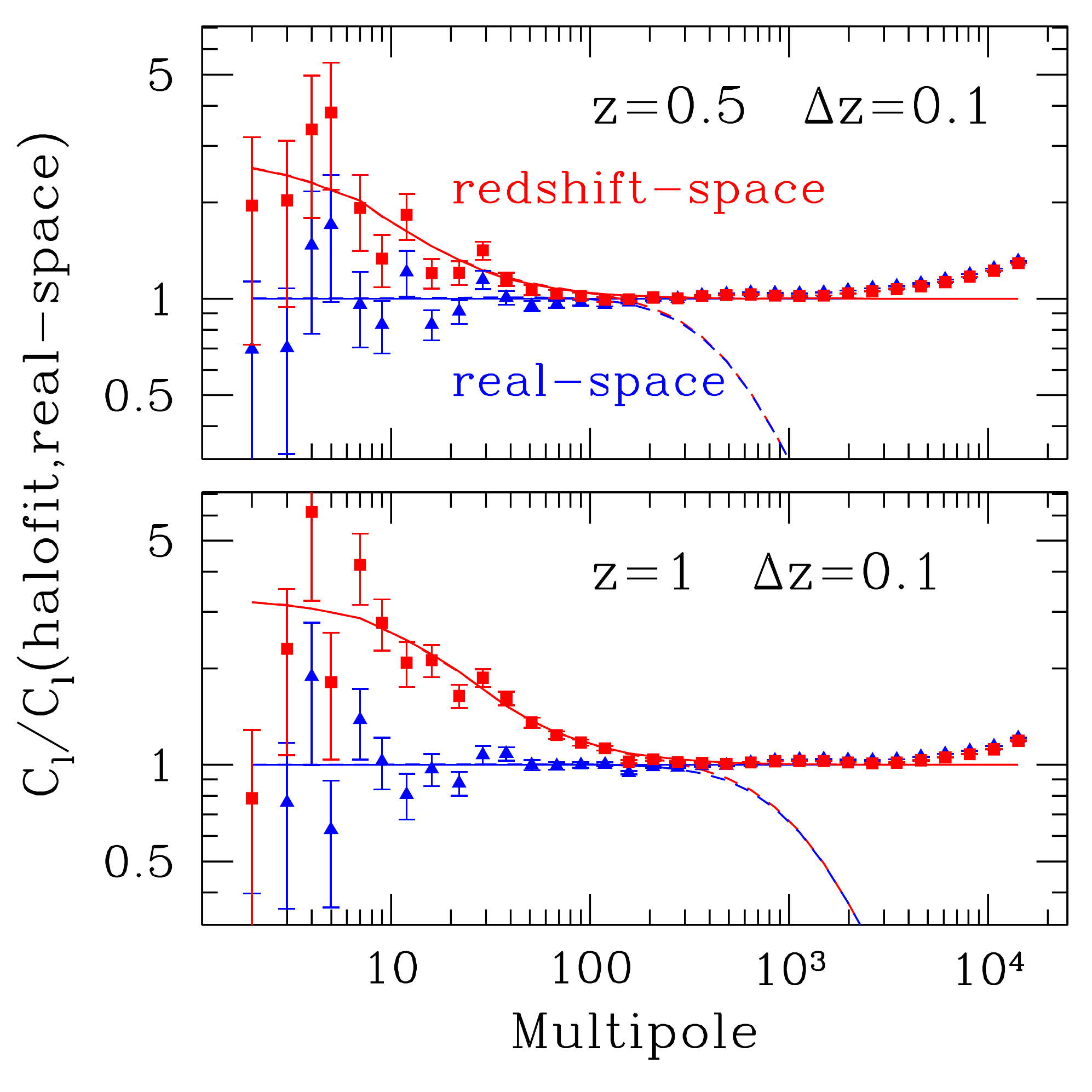}
\includegraphics[width=0.49\textwidth]{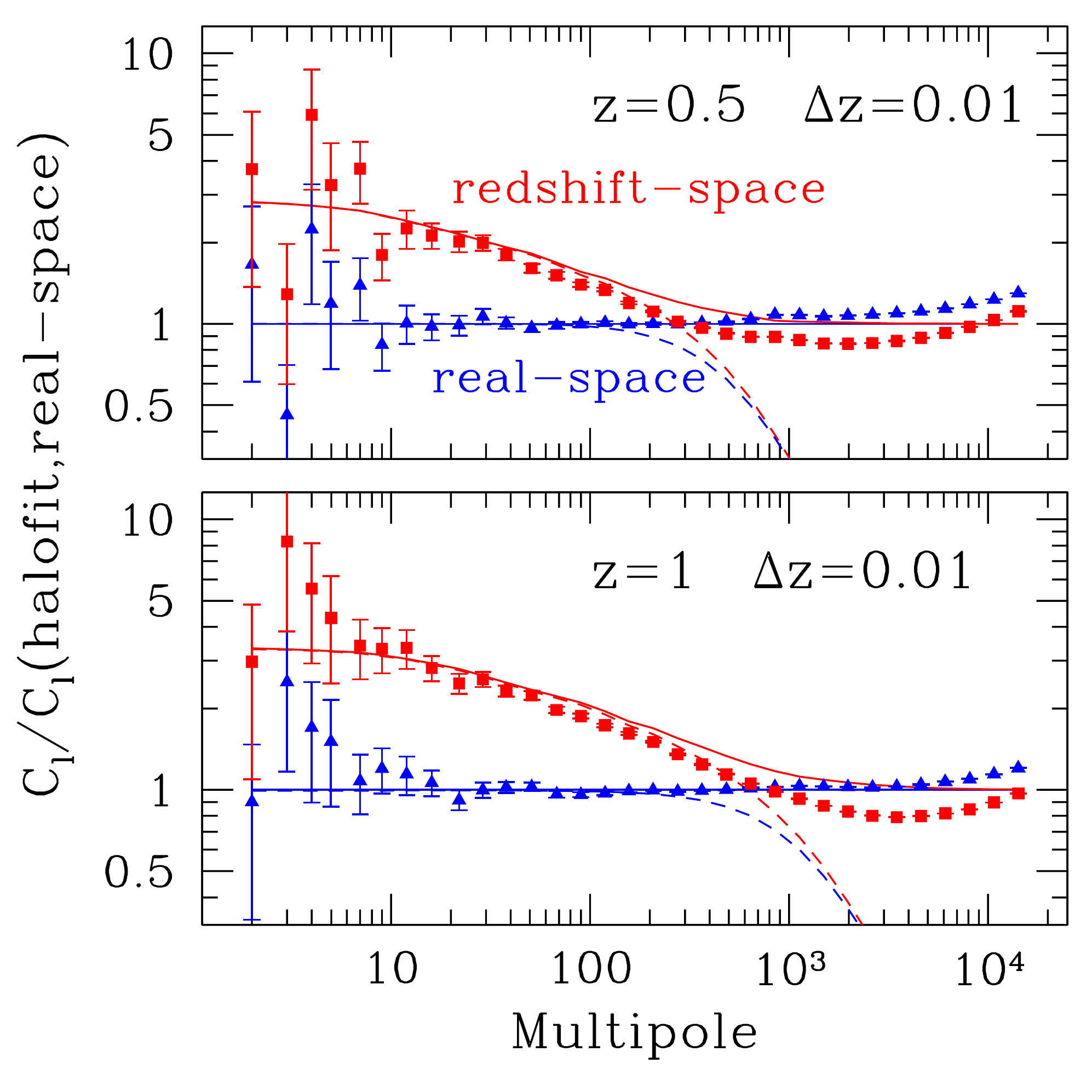} 
\caption{
Comparison between angular clustering in real
and redshift space. Left panels show results for $z=0.5,1$ for two broad redshift bins,
$\Delta z=0.1$.
Theory predictions are given in the linear (dashed) and non-linear
regime of gravitational clustering (Halofit, solid lines). Theory RSD
effects are only modeled in the linear regime (\ie Kaiser effect),
following the {\tt CAMB} implementation.
Symbols display measurements in the MICE-GC
simulation. Both theory and simulation results are normalized to
non-linear theory predictions in real-space (labeled
``halofit, real-space'' in the Figures).
Simulations show that RSD effects enhance the
large-scale (low-$\ell$) clustering relative to real-space,
by a factor of $\simeq 3$, in agreement with linear predictions (red
lines). Non-linear RSD tend to suppress power on small-scales. This is only
significant in 2D clustering for sufficiently narrow redshift bins, e.g, 
$\Delta z=0.01$ (see right panels).}
\label{fig:clrsd}
\end{center}
\end{figure*}

Let us start by comparing the angular power spectrum in 
lightcone dark-matter outputs
in real and redshift space from large linear scales (i.e low multipoles) to
small and non-linear scales (high-multipoles). In Figure \ref{fig:clrsd},  we show results for dark-matter sources in
the light-cone at mean redshifts, $z=0.5$ and $1$. In order to see the impact of
projection effects, we display results for two cases, broad $\Delta
z=0.1$ (left panels) and narrow $\Delta
z=0.01$ (right panels) redshift bins. 
Theory predictions, computed with {\tt CAMB sources}, are given in the linear and non-linear
regime for the growth of structure (Halofit, solid lines). 
However, theory RSD effects are only computed in the linear regime, the so-called
Kaiser effect.  

\begin{figure*}
\begin{center}
\includegraphics[width=0.49\textwidth]{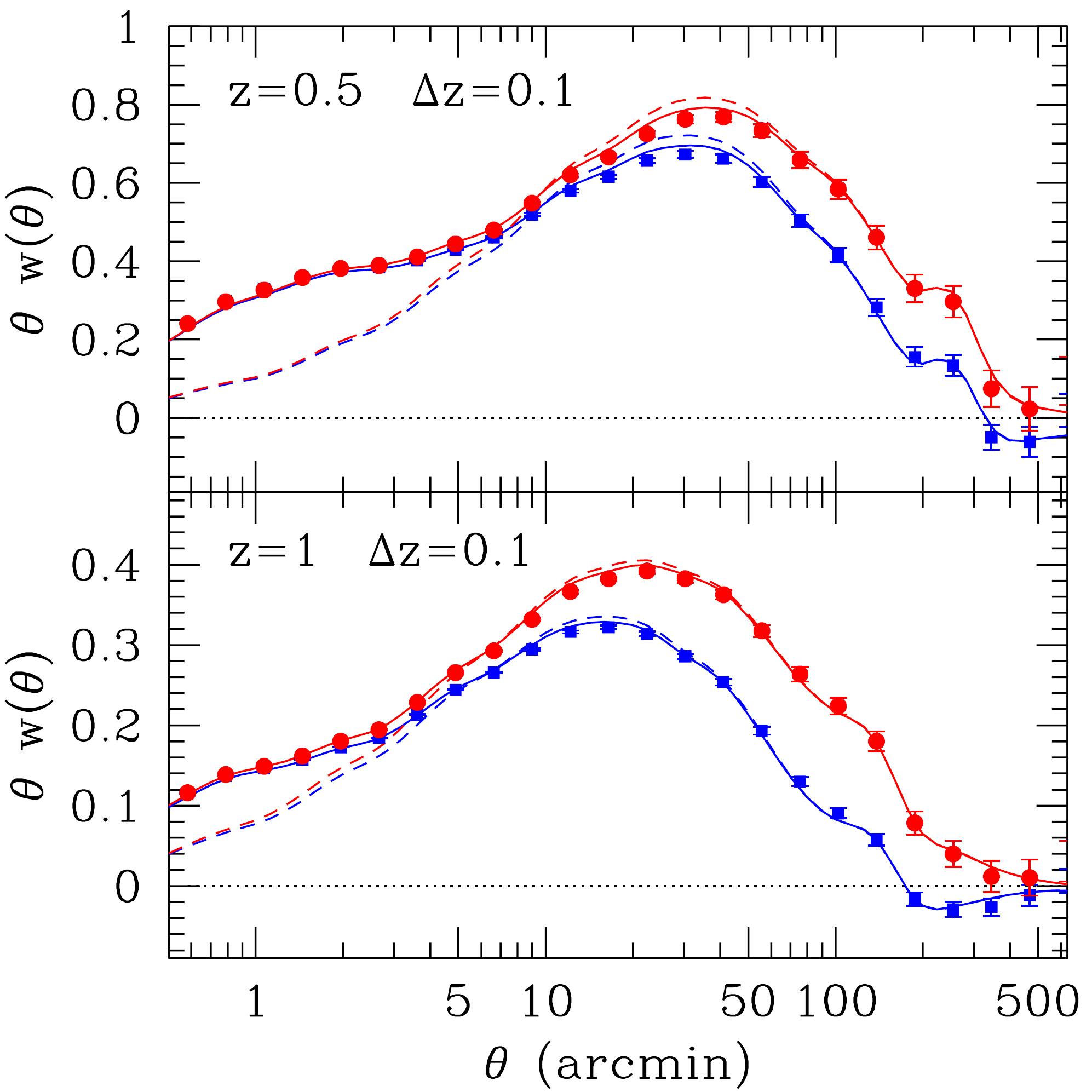}
\includegraphics[width=0.49\textwidth]{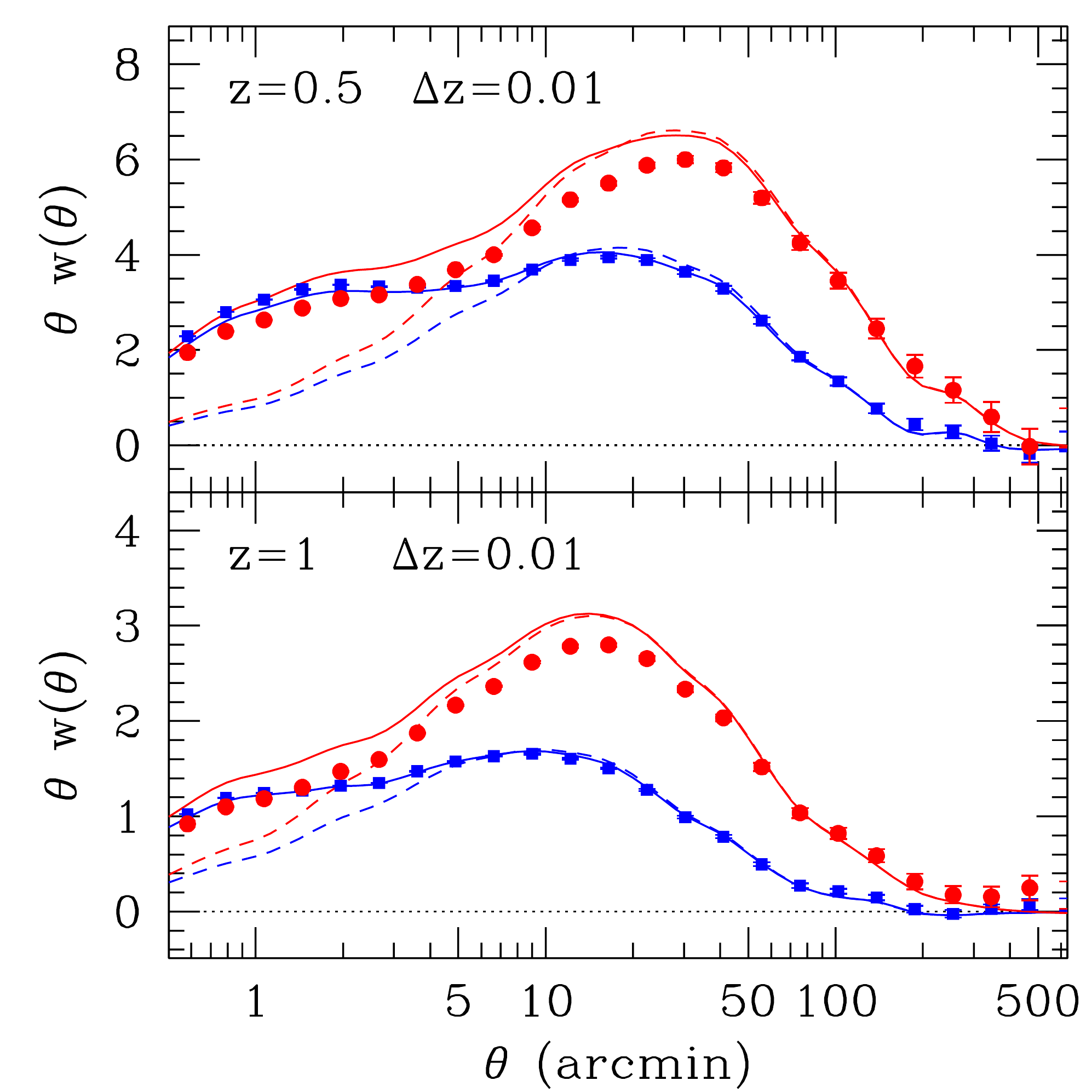} 
\caption{
Same as Figure~\ref{fig:clrsd} but for the angular 2PCF. The theory
predicted Kaiser effect is recovered in the simulation for $\theta \simgt 50-100$
arcmin, where coherent motions of dark-matter particles tend to enhance
clustering by a factor, $\simeq 3$. For narrow z-bins (right panels) non-linear RSD appear at $\theta \simlt
100$ arcmin, where the simulation (red symbols) deviates from
linear RSD theory  (red solid lines).}
\label{fig:wrsd}
\end{center}
\end{figure*}

In Fig.~\ref{fig:clrsd},  both theory and simulation results are
normalized to the
non-linear Halofit \citep{smith03} predictions in real-space (labeled
``halofit, real-space'' in plots).
Simulations show that RSD effects enhance the
large-scale (low-$\ell$) clustering relative to real-space,
by a factor of $\simeq 3$ (although it depends slightly on redshift),
in agreement with the Kaiser effect (e.g. \cite{2007MNRAS.378..852P}). 
This is also clearly seen in configuration space, see
Fig.~\ref{fig:wrsd}, where simulations accurately recover the Kaiser ``boost''  for
angular scales $\theta \simgt 50-100$ arcmin. 
In particular, the Kaiser boost also enhances the BAO feature on few degree
angular scales, an effect that has been studied in depth in the
context of upcoming photometric surveys \citep{crocce10,nock10}.

On the other hand, non-linear RSD caused by random motions in
virialized dark-matter halos
tend to suppress power on small-scales relative to 3D clustering in real-space. This is only
seen in 2D clustering for sufficiently narrow redshift bins, this is, when
projection effects do not cancel out completely random peculiar
motions along the line of sight.
We find a $\simeq 2 \%$ power suppression on non-linear growth
scales ($\ell \simgt \ell_{NL-growth} \simeq 300$) for $\Delta z
=0.1$, and we observe this relative suppression
grows roughly proportional to the inverse of the z-binwidth $\Delta z$, 
\beq
\left| \frac{\Cl(redshift-space) -\Cl(real-space)}{\Cl(real-space)}
\right|\simeq \frac{2\times 10^{-3}}{\Delta z}
\label{eq:rsdsupp}
\eeq
so that redshift-space angular clustering is $\simeq 20 \%$ lower than in
real-space for the narrow bin, $\Delta z =0.01$, in the non-linear regime.
It is interesting to see that non-linearities in RSD, as seen in the
simulation (see red symbols, as compared to linear RSD theory given by
the red lines), appear at much lower multipoles, 
$\ell_{NL-RSD} \simgt 30-50$, 
than those where gravitational growth enters the non-linear
regime, $\ell_{NL-growth} \simgt 500-10^3$,  where the lower (higher).
estimates correspond to $z=0.5$ and $1$ respectively.
In particular, we find that $\ell_{NL-RSD} \simeq
\ell_{NL-growth}/10$, for the bin widths studied. 
This is a consequence of the fact that velocities have more 
power on weakly nonlinear scales than densities, in addition to pairwise velocity
dispersion effects that enter in RSD (the so-called Finger-of-God effect).
For an order of magnitude estimate these effects enter through a
damping factor $\propto \exp(-f^2 k_z^2 \sigma_v^2)$ where $\sigma_v$
is the 1D linear velocity dispersion and $f$ the growth rate\footnote{This is discussed in detail in Sec. 6.2 of Paper
   II}.
If we set this to introduce a $10\%$ damping along the line-of-sight then $f^2
 \sigma^2_v \sim 0.1/k^2 \sim 0.1 \,r^2(z)/ \ell^2$. At $z=0.5$ where
 $\sigma_v \sim 5 \Mpc$ and $f \sim 0.7$
 this translates into $\ell_{NL-RSD} \sim 100-150$. Moreover we find a
 similar estimated value of $\ell_{NL-RSD}$ for $z=1$.
This is fully  compatible with our findings for narrow band binnings in
 Fig~\ref{fig:clrsd} (while broad-band binning erase to large degree this
 line-of-sight effects). 

Alternatively, non-linear RSD effects in the 2PCF $w(\theta)$ are significant
on scales $\theta_{NL-RSD} \simlt 100$ arcmin. In analogy to what we
found in harmonic space, this angular scale
is roughly an order of magnitude larger than the transition scale to non-linear
gravitational growth, $\theta_{NL-growth} \simeq \theta_{NL-RSD}/10 \simeq 10$ arcmin, as shown
in Figure~\ref{fig:wrsd}. The amplitude of non-linear RSD effects,
estimated as the difference between the linear RSD theory prediction
(red solid line) and the simulation measurement (red symbols) on
$\theta < \theta_{NL-RSD}$, is
$\simeq 2 \%$ on non-linear scales for the broad z-bin, $\Delta z =0.1$, and up to
$\simeq 5-15 \%$ for scales $\theta = 60 - 1$ arcmin respectively, for
the narrow z-bin. These amplitudes are
in agreement with the results found for the angular power spectra at
the highest multipoles (\ie non-linear scales), Fig.\ref{fig:clrsd}.

\begin{figure*}
\begin{center}
\includegraphics[width=0.33\textwidth]{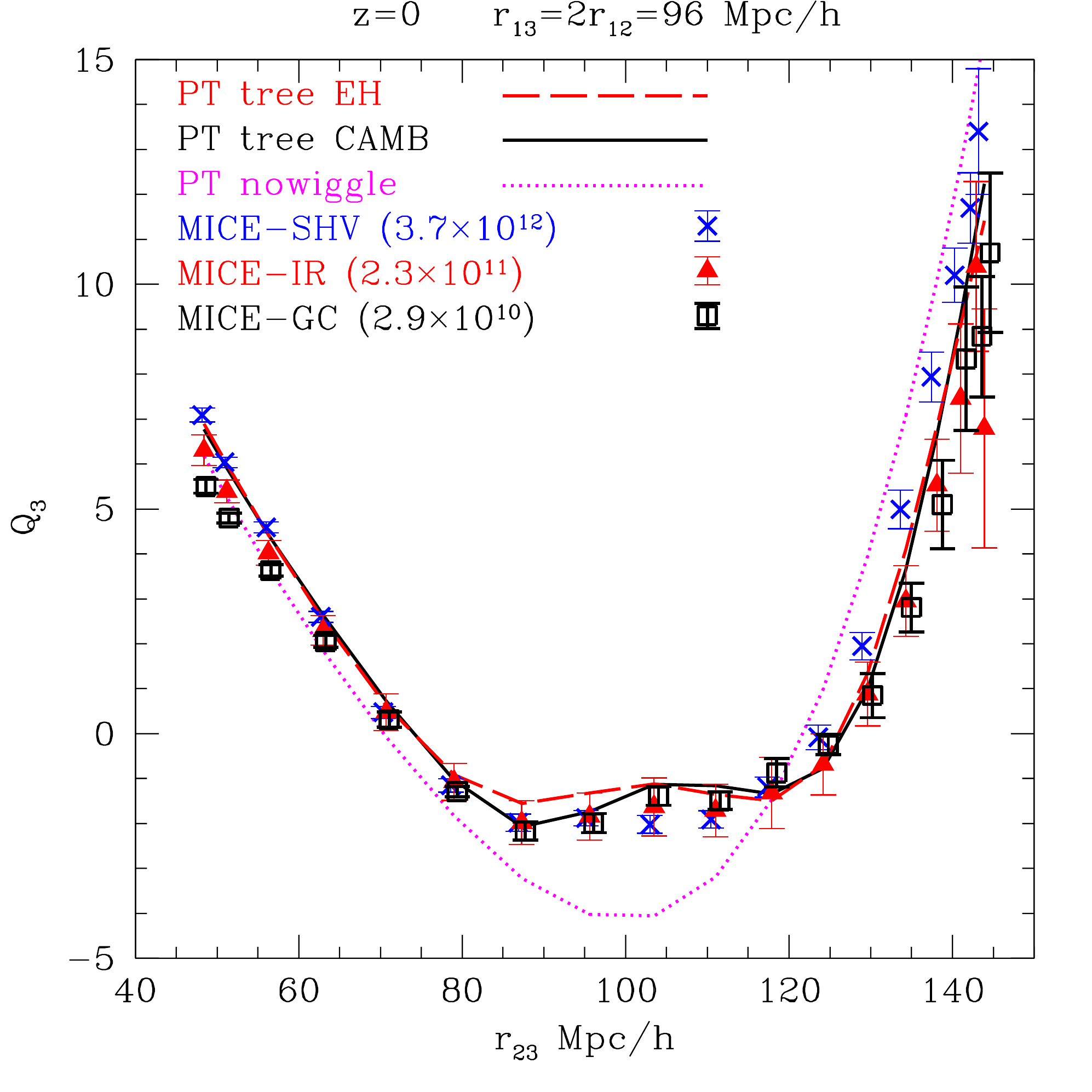} 
\includegraphics[width=0.33\textwidth]{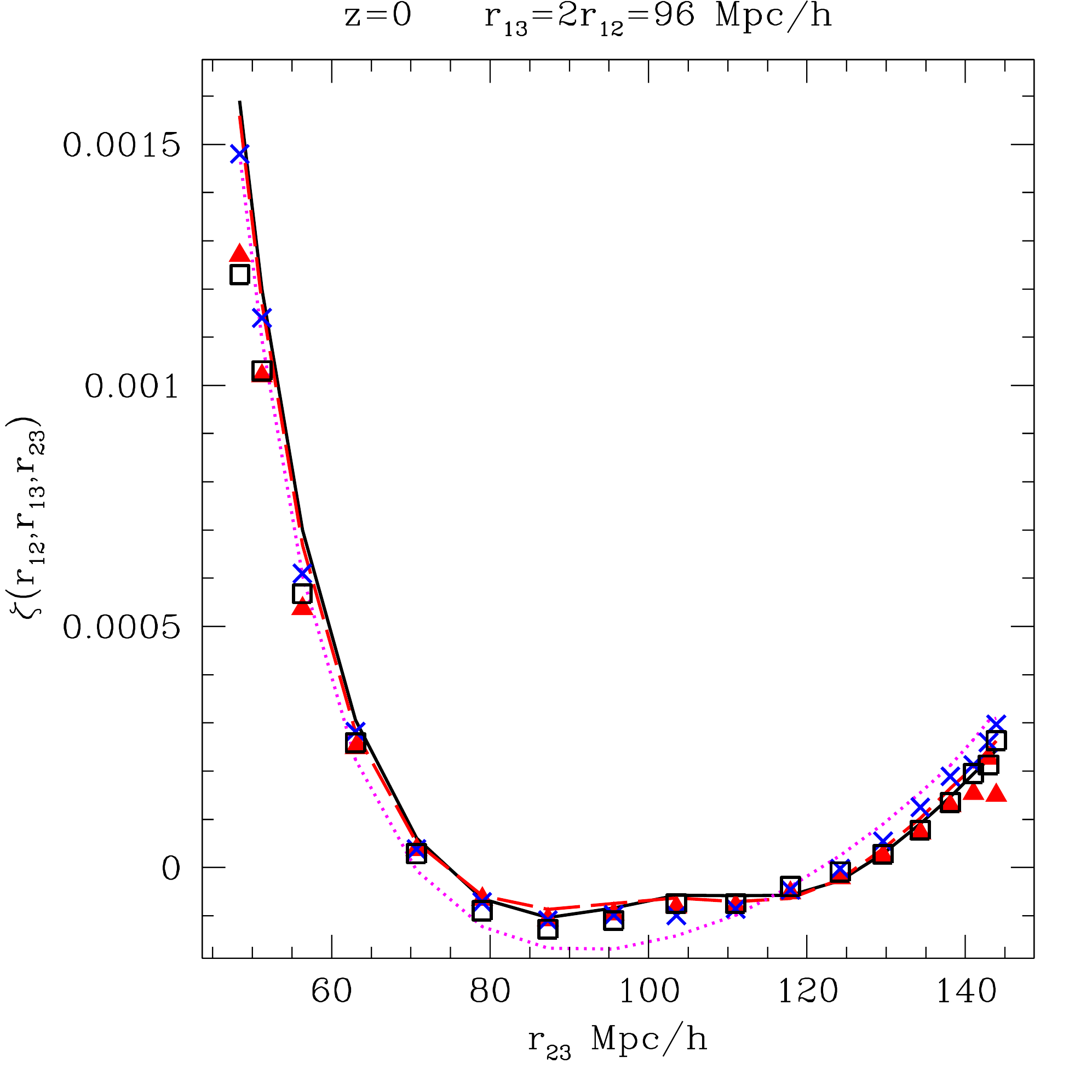} 
\includegraphics[width=0.33\textwidth]{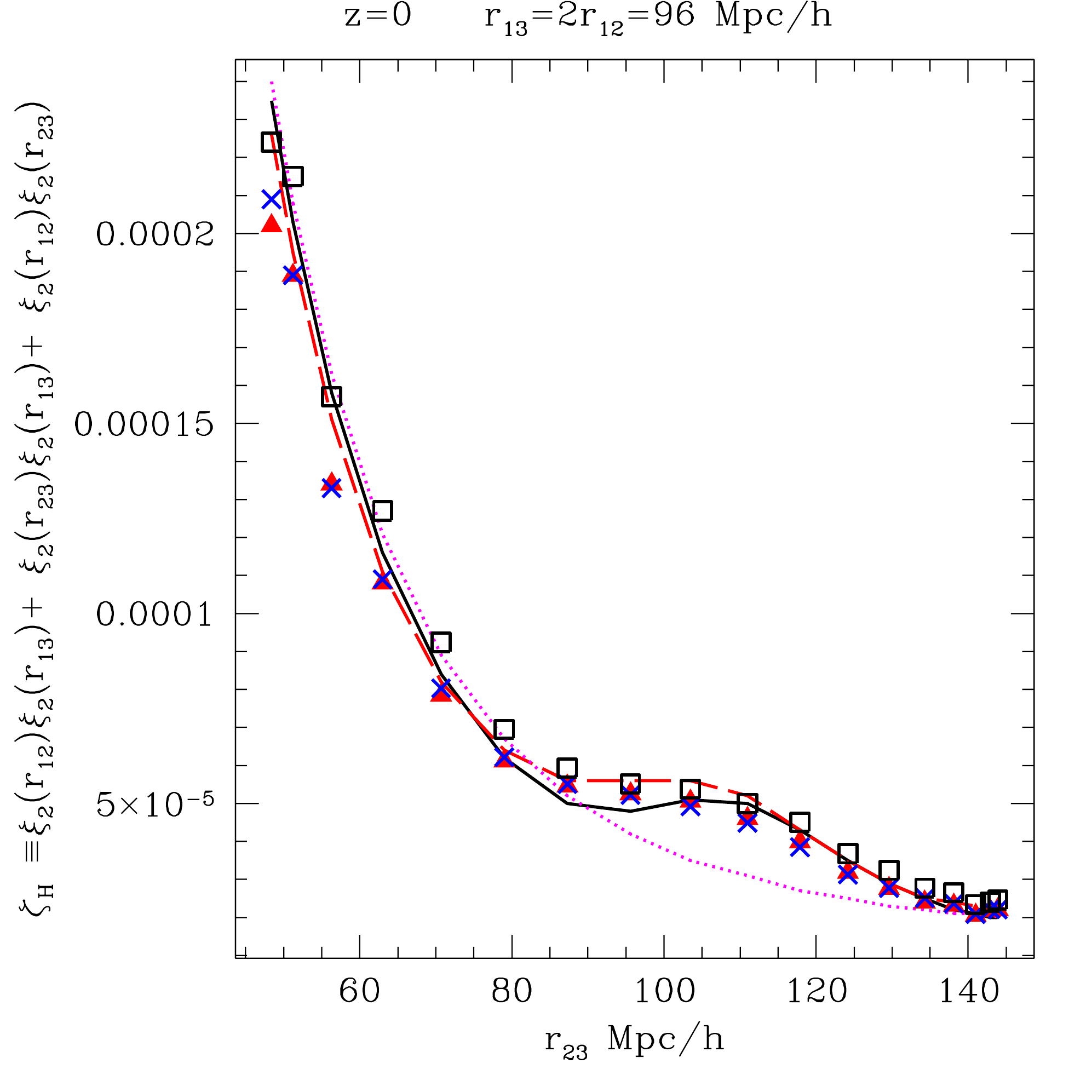} 
\caption{Reduced 3-point $Q_3(r_{12},r_{13},r_{23})$ (left panel), which is
the ratio, see Eq.~\ref{Q3def},
of the 3-point $\zeta(r_{12},r_{13},r_{23})$ (middle panel) to the hierarchical
2-point product $\zeta_H \equiv {\xi
  (r_{12})\xi(r_{23})+\xi(r_{12})\xi(r_{13})+\xi(r_{23})\xi(r_{13})}$
(right). This is  for z=0
and $r_{13}=2 r_{12}= 96$ Mp/h, as a function of $r_{23}$. MICE simulations 
of different resolutions (as labeled in the top panel) are compared to PT results with different
IC (transfer functions), including the case without BAO (no-wiggles).
The BAO peak can be clearly seen as a bump around $110 \Mpc$ in 
all panels. A lack of resolution reduces the power in the 2-point
$\zeta_H $, but increases slightly the anisotropy in the 3-point
$\zeta$.  Both effects  makes $Q_3$ significantly more
anisotropic  as we reduce the mass resolution in the MICE simulations. }
\label{fig:q3r96}
\end{center}
\end{figure*}

\section{Higher-order clustering: 3-point correlation function}
\label{sec:3pcf}

\subsection{Mass-resolution effects}

The 2 and 3-point correlation functions are defined, respectively, as
\bea
\xi(r_{12}) &=& \langle \delta(r_1) \delta(r_2) \rangle \\
\zeta(r_{12},r_{23},r_{13}) &=& \langle \delta(r_1) \delta(r_2) \delta(r_3) \rangle 
\label{2and3pfdef}
\eea
where $\delta(r)$ is the local density contrast at position $r$ smoothed over a given characteristic
R scale, and the 2-point function is the Fourier transform of the power
spectrum shown in \S\ref{sec:powerspectrum}. 
The reduced 3-point function $Q_3$ \citep{groth77} is defined as:
\bea
\label{Q3def}
Q_3 &=& \frac{\zeta(r_{12},r_{23},r_{13})}{\zeta_H(r_{12},r_{23},r_{13})} \\
\zeta_H  &\equiv&
{\xi
  (r_{12})\xi(r_{23})+\xi(r_{12})\xi(r_{13})+\xi(r_{23})\xi(r_{13})}, \nonumber
\eea
where we have introduced a definition for the "hierarchical" 3-point
function $\zeta_H$. Based on early galaxy measurements of $\zeta$,
the $Q_3$ parameter was thought to be roughly constant
as a function of triangle shape and scale, a result that is usually
referred to as the hierarchical scaling. It was later shown
\citep{peebles80, fry84} 
that one expects deviations from this scaling in the weakly non-linear
regime due to gravitational clustering starting from Gaussian initial conditions,
which enhances the filamentary structure and produce anisotropic values of
$Q_3$ and $\zeta$ as we change the shape and size of the triangles.

\begin{figure*}
\begin{center}
\includegraphics[width=0.45\textwidth]{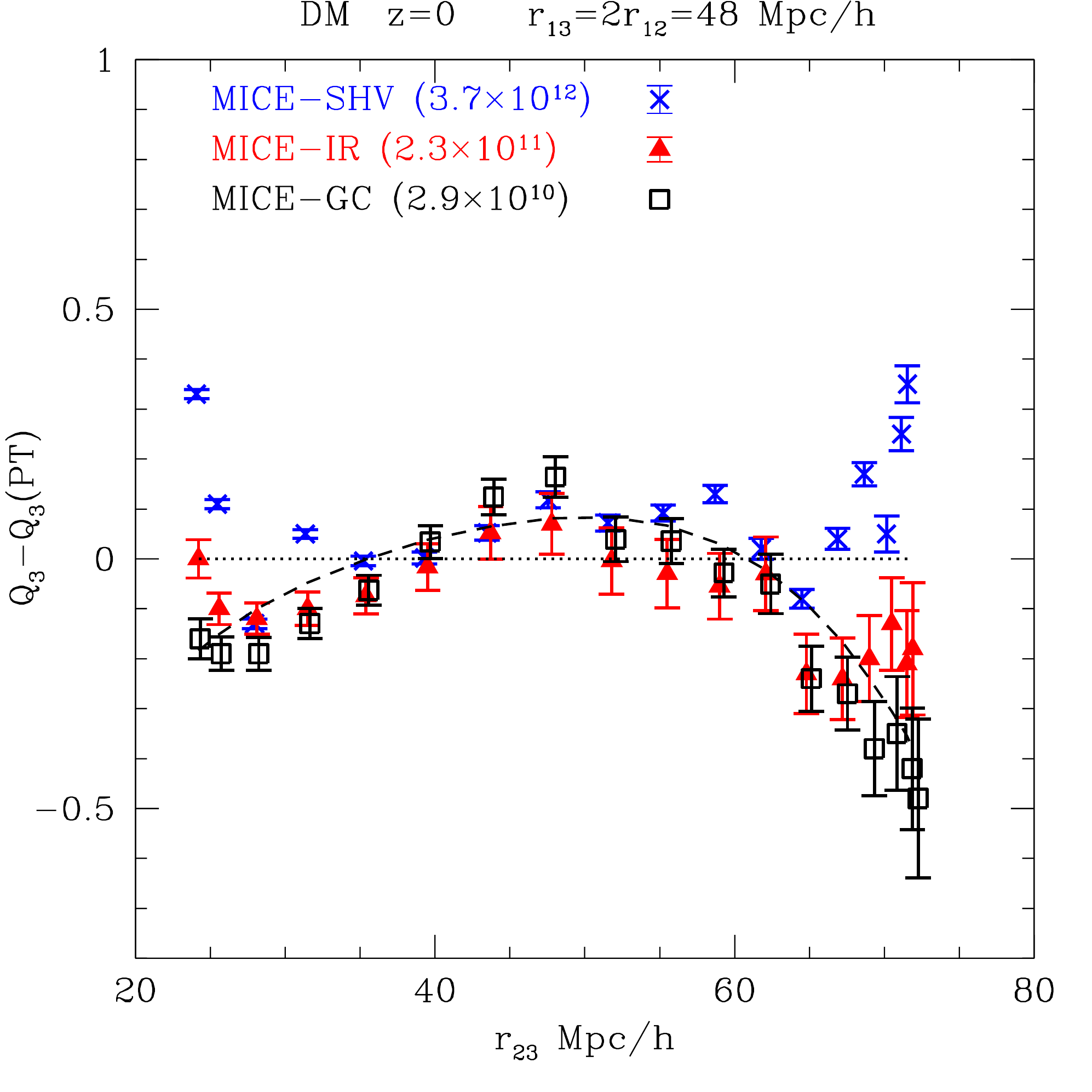} 
\includegraphics[width=0.45\textwidth]{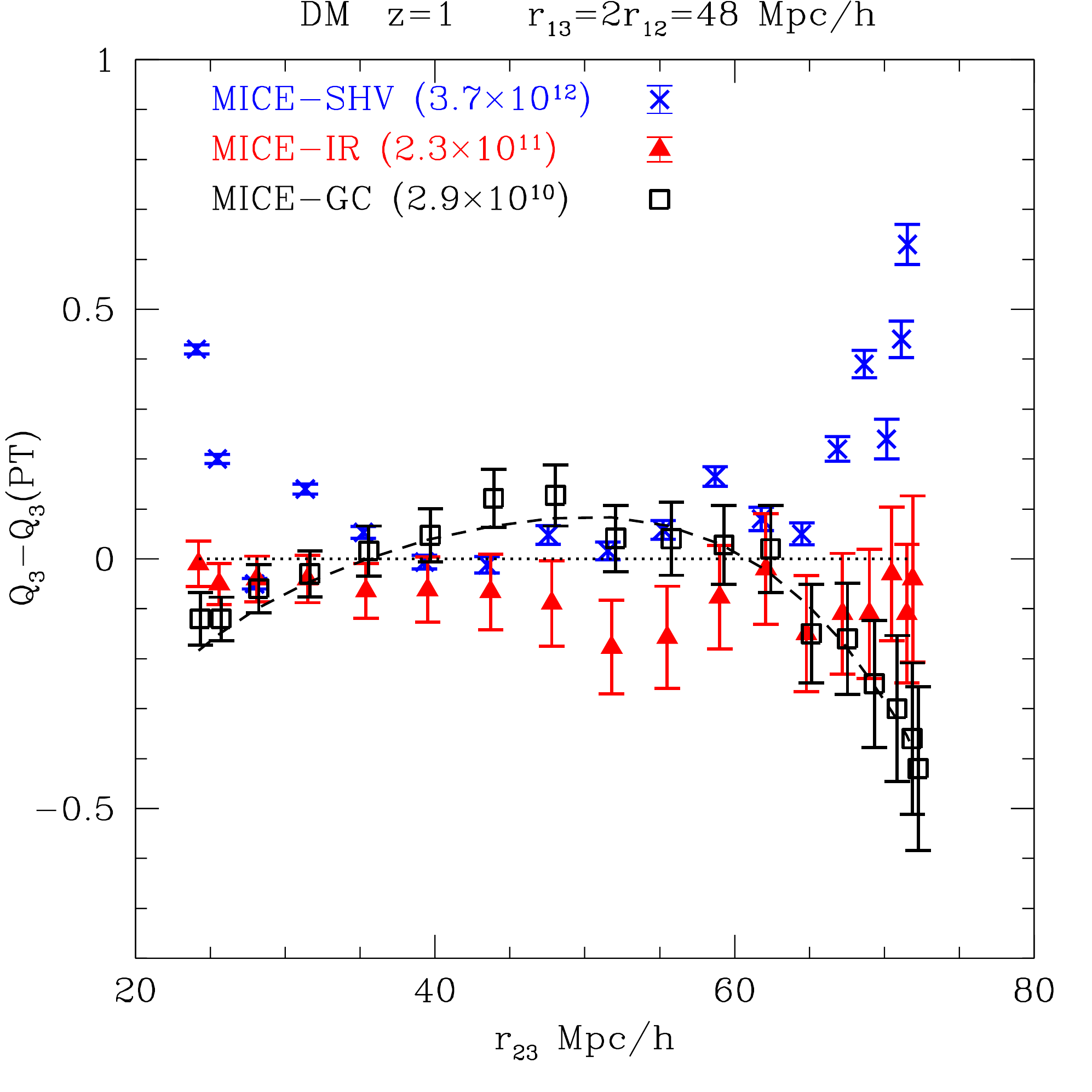} 
\caption{The difference between leading order perturbation theory
(PT) results and measurements in simulations with different resolutions.
In each case we use the corresponding transfer function to estimate
the PT results. This is for triangles with $r_{13}=2 r_{12}= 48$ Mp/h
and shown as a function of the remaining leg $r_{23}$. 
As the simulation evolves the different resolution converge to each
other (left, z=0),  but note how at earlier times (right, z=1), the
lower resolutions produce quite different results. The dashed lines
includes an additional tidal component that matches the high resolution
simulation (to guide the eye).}
\label{fig:dq3r48}
\end{center}
\end{figure*}

To  illustrate the dependence of  $Q_3$ on the triangle shape
we will fix two of the triangle sides ($r_{12},r_{13}$) and display
results as function of the third side  $r_{23}$. 
Figure \ref{fig:q3r96}
shows a comparison of $Q_3$ as measured from
the MICE simulations of different particle resolutions.
We also show the predictions from tree-level perturbation theory (PT,
from \cite{Barriga}) using different transfer functions in the
initial conditions (\ie {\tt CAMB} and EH) 
and the no-wiggle model of EH, i.e. without the BAO peak.
We have found very little change of $Q_3$  with redshift within the
errors, so these results are very similar for $0<z<1$
(see also Fig.~\ref{fig:q3r48r48} below).

We can see a small, but systematic and significant, dependence on the simulation
mass particle resolution that are comparable in order of magnitude 
to the deviations between simulations and PT and also the differences
between different transfer functions. The biggest discrepancy is with
the model without a BAO, indicating that the BAO can be clearly detected
in $Q_3$ in our simulations
(see \cite{2009MNRAS.399..801G} and references
therein for measurements of this effect). 
These results can only be achieved with the largest volumes
(and therefore at higher redshifts in the lightcone).

The discrepancies at the 2-point function level, i.e. in $\zeta_H$, are
comparable to what we show in previous sections and can be understood
using the halo model. The higher the resolution the lower mass halos
that we can resolve. This results in higher power on the smallest scales,
corresponding to the smallest mass halos that are resolved.
 For the 3-point function, the effect 
comes from the mode coupling which makes the clustering anisotropic.
This results in a characteristic anisotropic shape of $Q_3$ or $\zeta$
as a function of the triangle shape (which is given by a U-Shape 
as a function of $r_{23}$ in Fig.~\ref{fig:q3r96}). Non-linear dynamics
tend to reduce this  shape dependence and makes clustering more
isotropic, specially on the smaller scales
(e.g. see \cite{2002PhR...367....1B}). This could explain why a
better resolution of non-linear effects (which comes with higher
particle resolution)  results in slightly more
isotropic clustering, which in our case means slightly lower
amplitudes of $Q_3$ and $\zeta$ for elongated triangles, as
shown in  Fig.~\ref{fig:q3r96}.

Figure~\ref{fig:dq3r48} illustrates these arguments in more detail. Here we show the
differences between measurements and leading order perturbation
theory (PT) for triangles that are a factor of 2 smaller. This result
in smaller errors, that are useful to appreciate some trends.
We can see how MICE-GC, the higher resolution
simulation (squares), does not change much with redshift (compare right
to left panels) and shows deviations
from PT that are small but significant. These are caused by higher order
(loop) corrections in PT. As in the leading order contribution
to $Q_3$, these corrections also originate in the
non-linear mode coupling between scales.
To guide the eye, we match theses differences with an
 ad-hoc amplitude:
\beq
Q_3-Q_3^{PT} \simeq A  Q(\beta) +B 
\eeq
where  $Q(\beta)$ corresponds to the quudrupole in the mode coupling term from Euler's
equation (Eq.~(38) in \cite{2002PhR...367....1B}), i.e. gradients
of the velocity divergence in the direction of the flow. This is shown 
as dashed lines in Figure~\ref{fig:dq3r48}.  In contrast,  we can see in
Figure~\ref{fig:dq3r48} that the lower resolution simulations show
more evolution. At $z=0$ (left panel) all results tend to agree
(except the co-linear configurations for the lower resolution, which
are still converging). At $z=1$, the intermediate resolution (IR-MICE) show 
a significant difference in configuration, with a tendency of being
less sensitive to the configuration (shape of triangle) than the
higher resolution simulation. At higher redshifts, the resolution
effects become more important as non-linearities affect larger
scales than at lower redshifts. For the 2-point function
this results in lower clustering  (e.g. see discussion around Fig.5.).
For $Q_3$ this translates into smaller mode coupling and less
configuration dependence. For precision measurements and 
interpretation (of galaxy bias models, see paper II) these differences are significant
and affect resolutions ($\sim 10^{11}$) that are often used to study
the shape dependence in $Q_3$ and its biasing (e.g. see 
\cite{Manera} and references therein).

\begin{figure}
\begin{center}
\includegraphics[width=0.45\textwidth]{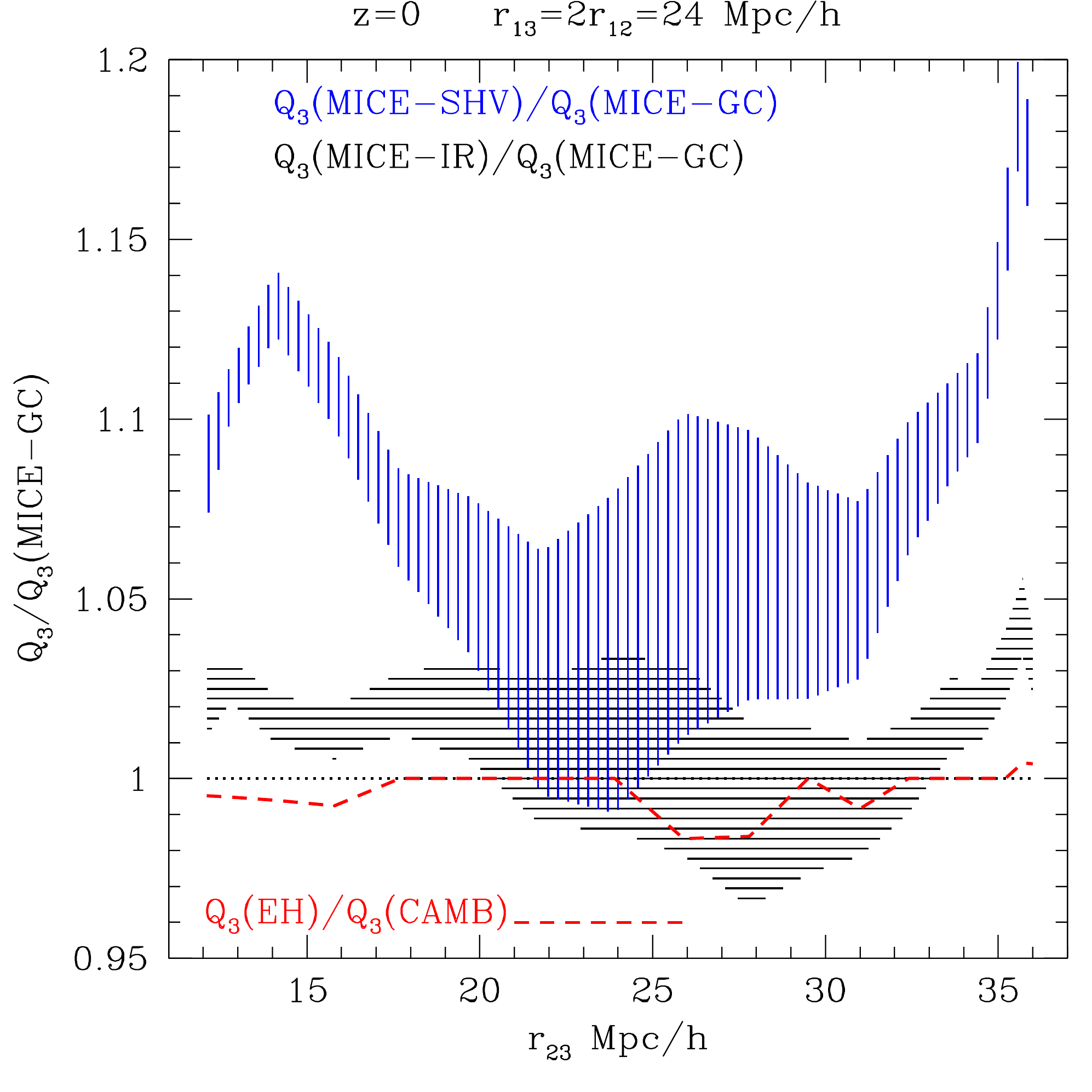} 
\caption{This is similar to Fig.\ref{fig:q3r96} 
but for smaller scales $r_{13}=2 r_{12}= 24$ Mp/h. Here we show the
ratio of the low resolution results with respect to the 
 one with highest resolution (MICE-GC). Deviations from unity are significant
given the errors (shaded regions) and do not seem to arise from small
differences in the transfer functions used (dashed lines).}
\label{fig:q3r24}
\end{center}
\end{figure}

Figure~\ref{fig:q3r24} shows these differences  more clearly at
$z=0$ and smaller scales ($r_{13}=2 r_{12}= 24 \Mpc$) as a ratio of $Q_3$ in
the lower resolution simulations (MICE-IR and MICE-SHV) with respect to the value of
$Q_3$ in the highest resolution (MICE-GC, $2.9\times 10^{10} \Msun$ particle mass). 
Error-bars are given by the shaded region. 
Overall deviations are around $5\%$ and as large as $10-15\%$ for
the lowest resolution run (MICE-SHV, $3.7\times 10^{12} \Msun$ particle mass).
For a particle mass of $2.3\times 10^{11}\Msun$, as in MICE-IR, deviations are around
2\% and always smaller than $5\%$. These deviations are significant,
given the error-bars, and can not be explained by the differences in
the transfer function shown in Fig.~\ref{fig:Pkmassresolution} which,
at these scales, are smaller than $1\%$ is $Q_3$ (\ie dashed line).

The observed trend of increasing mass-resolution effects as one samples
smaller (more non-linear) scales seems consistent with
previous analyses of the 3-point function on even smaller scales, \ie
$\sim$ 1 Mpc/h \citep{fosalba05}, where measurements on comparable
resolution simulations exhibit more configuration
dependence than the halo model predictions
(see bottom panels of their Fig.10).

\subsection{Realization effects}

\begin{figure}
\begin{center}
\includegraphics[width=0.47\textwidth]{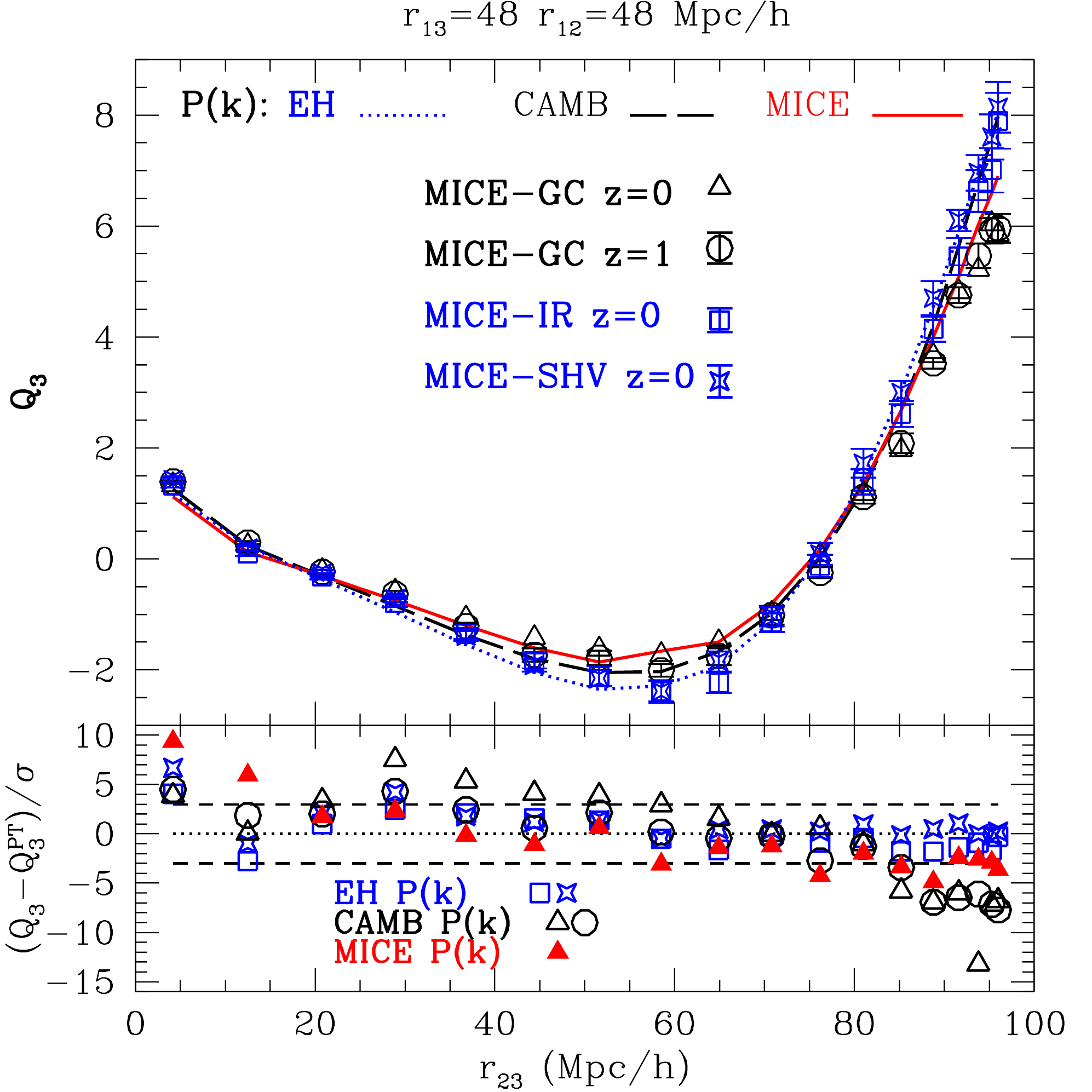} 
\caption{Comparison of $Q_3(r_{23})$  measurements (symbols)  for  $r_{13}=r_{12}= 48 \Mpc$
in different MICE simulations, with predictions (lines) based on
the initial transfer function $P(k)$ used in each simulation (EH for
IR and SHV, and {\tt CAMB} for GC). Measurements in MICE-IR (squares) and MICE-SHV (stars) agree well
with predictions from EH transfer function. MICE-GC measurements (triangles and circles)
are significantly lower than the  {\tt CAMB} predictions at the largest scales,
but give comparable results at z=0 and z=1.
The bottom panel shows the significance of the differences with
respect to the corresponding prediction.
The MICE-GC result compared to {\tt CAMB} (open circles and triangles) show more
than 5-$\sigma$ deviations at scales $r_{23}>90 \Mpc$. This discrepancy
reduces to 3-$\sigma$ when we use the
prediction estimated using the measured MICE-GC $P(k)$
(closed triangles).}
\label{fig:q3r48r48}
\end{center}
\end{figure}

In some particular triangular configurations, 
when we have very large scales $r \simeq 80-100 \Mpc$ but
small errors, we can detect some systematic differences
that seem to be caused by
the power spectrum realization. When using a single simulation,
the simulated spectrum is typically slightly different 
to input transfer at the largest scales  because of sampling variance.
This is illustrated in
Fig.~\ref{fig:q3r48r48} which compares configurations with $r_{12}=r_{13}=
48 \Mpc$ in MICE  simulations that uses EH and {\tt CAMB} transfer 
functions. On both the largest scales ($r_{23} \simeq 90-100 
\Mpc$) and intermediate scales ($r_{23} \simeq 40-70 \Mpc$) the values of
$Q_3$ are significantly different in the two simulations.
The differences at intermediate scales can be understood because of the
difference in the initial transfer function of the simulations. When we compare the results to
the corresponding PT predictions (in the bottom panel) we find no
significant deviations (within 3-$\sigma$ errors) between measurements
and (tree-level) PT results on this intermediate scales,  also indicating that particle
resolution does not seem so important for these configurations, given
the errors. Note that the errors in MICE-GC and MICE-IR are from
Jack-knife (JK) resampling, while the error in MICE-SHV are obtained by
dividing the very big box into 27 subsamples.

 On the largest scales, the differences between the EH and
{\tt CAMB} transfer functions are small (\ie compare dotted line to dashed
line, which lie on top of each other for $r_{23} >90 \Mpc$), but
the  MICE-GC simulation (circles and triangles) seem significantly
lower than the MICE-IR or MICE-SHV results. We find very similar results at
other redshifts. The differences are not caused by
the binning in the scales defining the sides of the triangles 
$(r_{12},r_{13},r_{23})$ and the effect does
not seem to be caused  by resolution (given the agreement 
within errors with the corresponding predictions at
lower scales and the agreement on all scales between MICE-IR and
MICE-SHV).
 These differences seem to originate in sampling variance in the
realization of the initial transfer function of the MICE-GC
simulation on the smallest k-modes.\footnote{This variance is not
  captured by our $Q_3$ JK error-bars as it is based on sampling the
spatial variations within the one single realization of the {\tt CAMB}
power spectrum.}
 This is hinted when comparing the open and closed (red)
triangles in the bottom panel. Here the open triangles
compare MICE-GC (z=0) measurements 
to the {\tt CAMB} prediction (which is the input to the MICE-GC
simulation), 
while closed triangles compare the same measurements to the
predictions based on the actual measured MICE-GC power spectrum. 
This is different from the {\tt CAMB} predictions because of sampling
variance in the MICE-GC realization of the {\tt CAMB} transfer function. 
While deviations with the {\tt CAMB} $P(k)$
predictions are quite significant (between 5 and 10 $\sigma$), deviations
with the MICE-GC $P(k)$ predictions are lower (within 3 $\sigma$). 
The values at z=1 (squares) show a similar trend,
indicating that this is not just an evolution effect. This figure
also illustrates how $Q_3$ does not evolve much with redshift (compare
circles, z=1,  and open triangles, z=0) and 
that PT predictions are more accurate for intermediate scales than
for small  or large scales: \ie it fails more for triangles that are
collapsed (see also \cite{2002PhR...367....1B}).
As discussed above, the lower resolution simulations (squares and stars) seem
to agree better with PT results because of the artificial suppression of 
non-linearities (\ie due to poor mass resolution).

 Once we use the right shape of the power spectrum, the
measurements seem to match predictions within 3-$\sigma$ (dashed 
lines in bottom panel), although the agreement is even better for
triangles that are less collapsed. In general, we do not expect
tree-level PT to match the simulations perfectly and
some deviations  are expected due to higher-order (loop) corrections 
 (see \cite{2002PhR...367....1B}).


\section{Conclusions}   
\label{sec:conclusions}   

We have presented one of the largest Nbody runs
completed to date: the MICE Grand Challenge Lightcone simulation
(MICE-GC), containing about 70 billion particles in a 3 Gpc/h periodic
box. 
This combination of large volume and fine mass resolution 
allows to resolve the growth of structure form the largest
(linear) cosmological scales down to very small (tens of kpc's) scales,
well within the non-linear regime.
Therefore, the MICE-GC presents multiple potential applications to study the
clustering and lensing properties of dark-matter and galaxies
that can be confronted with observations from upcoming galaxy surveys. 

Furthermore we have populated halos with galaxies using a hybrid HOD
and HAM scheme and studied their clustering and lensing properties. 
Clustering results from the halo and galaxy catalogs are presented in
an accompanying paper (Paper II, \citet{MICE2}), whereas the
all-sky lensing maps and galaxy lensing properties modeled with this
new mock are discussed in Paper III \citep{MICE3}. 
Further details about the galaxy assignment method implemented to
build the MICE-GC mock galaxy catalog will be presented in 
forthcoming papers (Carretero et al. 2014, Castander et al. 2014).

We make a first public data release of the MICE-GC galaxy mock,
{\tt MICECAT 1.0}, through a dedicated web-portal for simulations, hosted by
the Port d'Informacio Cientifica (PIC): {\texttt http://cosmohub.pic.es},
where detailed information on the data provided can be found.

In this first paper of the series (Paper I), we have discussed a basic
validation and the applications for the dark-matter
comoving and lightcone outputs, using 2D and 3D statistics. 
Throughout the paper we have investigated how mass-resolution effects
impact dark-matter clustering in the non-linear regime.
In other words, we have discussed how much our N-body measurements on
small scales might be limited by the particle mass used.
Although a proper systematic study based on an ensemble of
simulations will be presented elsewhere, we have obtained a series of
results and conclusions from the MICE-GC run alone that we summarize
below.

The main findings of this paper are as follows:
\begin{itemize}


\item{Using MICE-GC we can measure the BAO pattern in the 3D
    dark-matter power spectrum with high precision. We  find very good
    agreement with Renormalized Perturbation Theory (RPT) predictions
    at two-loops and $k \lesssim 0.2-0.3 \kvecMpc$, see Fig.~\ref{fig:BAO}. There is
    also a good match comparing to state-of-the-art numerical fits,
    although we find a slight excess broad-band power ($\lesssim 2\%$) with respect to the extended Coyote emulator (Heitmann et al. 2013) and the revised Halofit (Takahashi et al. 2012) at $z>0$. We note that these differences appear larger than our statistical error bars on those scales (see Figs.~\ref{fig:BAO} and \ref{fig:Coyote}).}

\item{ 
Detailed comparison across redshift evolution and nonlinear scales 
between the power spectrum measured in MICE-GC and
recent numerical fits shows that they agree well, but the latter seem
to slightly over-predict the power on BAO scales, specially at $z>0$ (see Fig.\ref{fig:Coyote}).
This is more clearly seen beyond BAO scales ($k> 0.3 \Mpc$) for the
recently revised Halofit fitting formula where the effect reaches
$5\%-8\%$ depending on redshift.}

\item{
By comparing the 3D power spectrum of MICE-GC to an order of magnitude
lower resolution run with the same cosmology, the MICE-IR, we conclude that
the change in clustering power produced by mass resolution effects 
increase with decreasing scale and grow with
redshift. Resolution effects are within 2 $\%$ for $k<0.2 \kvecMpc$
for $z<1$, but can be as large as 5 $\%$ for $k\simeq 1 \kvecMpc$ and
$z=1$ , as shown in Fig.~\ref{fig:Pkmassresolution}}.

\item{
We have also produced lightcone outputs of the MICE-GC run,
following the approach presented in \cite{fosalba08}.
As shown in Fig.~\ref{fig:clmassres}, the analysis of angular
clustering of the projected dark-matter in the lightcone yields consistent qualitative
results: the MICE-GC measures a $5\%$ excess power for  multipoles $\ell \sim 10^3$ with respect to
the lower-resolution run. But this grows to $20-30\%$ excess for multipoles
$\ell > 10^3$, which correspond to $k\simgt 0.75$ at $z=0.5$, and $k\simgt
0.5$ for $z=1$ in the Limber or small-angle limit. 
This confirms that the observed power excess is largely due to mass
resolution effects.
Moreover the impact of such resolution effect is
even larger for higher redshift, as shown in Fig.~\ref{fig:clmassres}.
This is consistent with what we found from the 3D power
spectrum analysis. Similar trends are found
in the angular 2-point function, as shown in Fig.~\ref{fig:wmassres}.}

\item{
Comparison of angular power spectra with recent numerical fits shows
that the MICE-GC measurements show an $5-10\%$ excess relative to
 Halofit \citep{smith03} in the range $10^3<\ell<10^4$, 
and a $5-10\%$ deficit with respect to the
revised Halofit by \cite{takahashi12}.}

\item{
The modeling of the RSD effects in angular clustering, investigated in \S\ref{sec:clrsd},
concludes that MICE-GC recovers the Kaiser effect on large scales, \ie
a boost of power of order $\simeq 3$ relative to the real-space
counterpart.
On the other hand, non-linear RSD caused by random motions in
virialized dark-matter halos
tend to suppress power on small-scales relative to 3D clustering in
real-space. This is only seen in 2D clustering for sufficiently narrow
redshift bins, and the amplitude of the effect roughly scales with the
inverse of the redshift binwidth (see Eq.~\ref{eq:rsdsupp}).
In particular, non-linearities in RSD, as seen in the
simulation (see Fig.~\ref{fig:clrsd}) appear at much lower multipoles 
than those where gravitational growth enters the non-linear
regime. We find that $\ell_{NL-RSD} \simeq \ell_{NL-growth}/10$, and
similar results hold for the angular 2PCF, \ie $\theta_{NL-growth} \simeq \theta_{NL-RSD}/10 \simeq 10$ arcmin.}

\item{ 
For the 3-point function, we find consistent
results with those of the 2-point statistics. Higher mass resolution runs resolve
better the small-scale power which tends to flatten out the reduced 3-point
function, $Q_3$ with respect to PT (tree-level) expectation. This
effect seems to get worse at higher redshifts, as shown in Fig.~\ref{fig:dq3r48}.
We show how the BAO signal can be detected in $Q_3$, as indicated by 
measurements by \cite{2009MNRAS.399..801G}. On BAO scales errors are
quite large and resolution effects are not very significant. On small
scales ($12-24 \Mpc$),  Fig.\ref{fig:q3r24} shows how the lower
resolution simulations, tend to estimate a more anisotropic 3-point function
than the higher resolution ones, at the $5\%$ level for MICE-IR and up to $15 \%$ for
the MICE-SHV run (which has about 100 times larger particle mass than
MICE-GC). On intermediate scales, the mass resolution  produces
differences that are systematic and significant. We have
learned here  that $Q_3$ is quite sensitive to the
actual shape of the power spectrum but this can be degenerate with
resolution effects for larger volumes, such as the ones in MICE simulations.}

 \end{itemize}

In summary this new large-volume simulation is shown to describe
accurately dark-matter clustering statistics in a wide dynamic
range. From the linear to the highly non-linear regime of
gravitational clustering, thanks to its adequate combination of
large-volume and good mass resolution. 
The resulting galaxy mock {\tt MICECAT v1.0},
described in the accompanying Papers II and III of this series, is
made publicly available with the hope that it will be used as a
powerful tool to develop and exploit the high quality data 
that is expected from the new generation of astronomical surveys.

\section*{Acknowledgments} 
We would like to thank Volker Springel for useful discussions
regarding Gadget2, Antony Lewis and Ruth Pearson for support with the
{\tt CAMB sources} code, Eric Hivon for help with Healpix and
discussions about spherical harmonic analysis, and the Barcelona
Supercomputing Center (BSC) support staff, specially David Vicente, for their continued support to
develop this Grand Challenge simulation. Special thanks to Santi
Serrano for his great help with the MICE website, the simulation
visualizations, including Figure 1 in this paper, and for enlightening
discussions that led to the development of the simulations web-portal.
We are greatly indebted to Jorge Carretero, Christian Neissner, Davide
Piscia, Santi Serrano and Pau Tellada for their development of the
CosmoHub web-portal.
We acknowledge support from the MareNostrum supercomputer (BSC-CNS,
www.bsc.es), through grants AECT-2008-1-0009, 2008-2-0011  and 2008-3-0010,
2009-1-0009, 2009-2-0013, 2009-3-0013,  and Port d'Informaci\'o Cient\'ifica (www.pic.es) where the
simulations were ran and stored, respectively, and 
the use of the Gadget-2 code to implement the N-body (www.mpa-garching.mpg.de/gadget).
Funding for this project was partially provided the European
Commission Marie Curie Initial Training Network CosmoComp
(PITN-GA-2009 238356),
the Spanish Ministerio de Ciencia e Innovacion (MICINN),
projects 200850I176, AYA-2009-13936, AYA-2012-39620, AYA-2012-39559, 
Consolider-Ingenio CSD2007-00060 and research project SGR-1398
from Generalitat de Catalunya. MC acknowledges support from the Ram{\'o}n
y Cajal MICINN program.

\bibliography{mn.bib}

\end{document}